\documentclass{LMCS}

\def\doi{8 (1:19) 2012}
\lmcsheading%
{\doi}
{1--44}
{}
{}
{Mar.~\phantom09, 2011}
{Mar.~\phantom06, 2012}
{}

\usepackage{amssymb,amsmath}
\usepackage{enumerate}
\usepackage{hyperref}
\usepackage{latexsym}
\usepackage{exscale}

\def\vec#1{\mathchoice{\mbox{\boldmath$\displaystyle#1$}}
{\mbox{\boldmath$\textstyle#1$}}
{\mbox{\boldmath$\scriptstyle#1$}}
{\mbox{\boldmath$\scriptscriptstyle#1$}}}

\newlength{\hilflh}
\newcommand{\hilfminus}[1]{
  \settowidth{\hilflh}{$#1-$}\mbox{$#1-\hspace{-0.5\hilflh}
  \makebox[0pt]{\raisebox{0.24\hilflh}{$#1\cdot$}}\hspace{0.5\hilflh}$}}
\newcommand{\minusp}{\mathbin{\mathchoice {\hilfminus{\displaystyle}}
  {\hilfminus{\textstyle}}{\hilfminus{\scriptstyle}}
  {\hilfminus{\scriptscriptstyle}}}}

\newcommand{\Fin}{\hfill$\Diamond$}

\newcommand{\imp}{\Rightarrow}
\newcommand{\aeq}{\Leftrightarrow}
\newcommand{\ssq}{\subseteq}
\newcommand{\fa}[1]{\forall #1\:\,}
\newcommand{\ex}[1]{\exists #1\:\,}
\newcommand{\andsp}{\;\&\;}

\newcommand{\al}{\alpha}
\newcommand{\be}{\beta}
\newcommand{\ga}{\gamma}
\newcommand{\de}{\delta}

\newcommand{\epsn}{\varepsilon_0}
\newcommand{\lam}{\lambda}
\newcommand{\om}{\omega}
\newcommand{\si}{\sigma}

\newcommand{\GT}{\mathrm{T}}
\newcommand{\GTn}{\mathrm{T}_{\!n}}
\newcommand{\CL}{\mathrm{CL}}
\newcommand{\GTlam}{\GT^\lam}
\newcommand{\GTCL}{\GT^\CL}
\newcommand{\FV}{\mathrm{FV}}
\newcommand{\BV}{\mathrm{BV}}

\newcommand{\rhdal}{\rhd_\al}
\newcommand{\lbr}{\lam\be\R}
\newcommand{\lbrpr}{\lam\be\R^\prime}

\newcommand{\pathP}{{\mathcal P}}

\newcommand{\treeR}{{\mathcal R}}
\newcommand{\treeS}{{\mathcal S}}

\newcommand{\lv}{\mathrm{lv}}
\newcommand{\Lv}{\mathrm{L}}
\newcommand{\Rv}{\mathrm{R}}
\newcommand{\DT}{\mathrm{D}_\GT}
\newcommand{\DTn}{\mathrm{D}_{\GTn}}
\newcommand{\DTF}{\mathrm{D}_F}
\newcommand{\lh}{\mathrm{lh}}
\newcommand{\wt}{\mathrm{w}}
\newcommand{\T}{\mathcal{T}}
\newcommand{\Tn}{\mathcal{T}_{n}}
\newcommand{\Tpr}{\T^\prime}
\newcommand{\Var}{\mathcal{V}}
\newcommand{\ot}{\mathcal{O}}
\newcommand{\otvar}{\ot_\Var}
\newcommand{\otlom}{\mathcal{O}^{<\om}}
\newcommand{\no}{\mathrm{no}}

\newcommand{\Nat}{{\rm I}\!{\rm N}}

\newcommand{\R}{\mathsf{R}}

\newcommand{\Rt}{{\mathsf{R}}_\tau}
\newcommand{\N}{\mathsf{0}}
\newcommand{\Suc}{\mathsf{S}}
\newcommand{\D}{\mathsf{D}}
\newcommand{\Dt}{{\mathsf{D}}_\tau}

\newcommand{\xsi}{{X^\si}}

\newcommand{\yrho}{{Y^\rho}}

\newcommand{\xsinull}{{x^\si_0}}
\newcommand{\xsii}{{x^\si_i}}

\newcommand{\yrhoi}{{y^\rho_i}}

\newcommand{\veca}{{\vec{a}}}
\newcommand{\vecb}{{\vec{b}}}
\newcommand{\vecc}{{\vec{c}}}
\newcommand{\vecd}{{\vec{d}}}
\newcommand{\vece}{{\vec{e}}}
\newcommand{\vecf}{{\vec{f}}}
\newcommand{\vecg}{{\vec{g}}}
\newcommand{\vech}{{\vec{h}}}
\newcommand{\vect}{{\vec{t}}}
\newcommand{\vecx}{{\vec{x}}}
\newcommand{\vecxsi}{{\vec{x}^\si}}
\newcommand{\vecy}{{\vec{y}}}

\newcommand{\vecal}{{\vec{\al}}}

\newcommand{\B}{{\mathcal B}}
\newcommand{\Bn}{{\mathcal B}_0}
\newcommand{\Be}{{\mathcal B}_1}
\newcommand{\Bi}{{\mathcal B}_i}

\newcommand{\Bip}{{\mathcal B}_{i+1}}

\newcommand{\C}{{\mathcal C}}
\newcommand{\Ci}{\C_i}
\newcommand{\Cx}{{\mathcal C}^\vecx}
\newcommand{\Cy}{{\mathcal C}^\vecy}
\newcommand{\Cxn}{\Cx_0}
\newcommand{\Cxe}{\Cx_1}
\newcommand{\Cxi}{\Cx_i}

\newcommand{\Cxk}{\Cx_k}
\newcommand{\Cxip}{\Cx_{i+1}}

\newcommand{\psiterm}[2]{\psi(\om \cdot #1 + #2)}

\newcommand{\bop}{\Box}
\newcommand{\bopd}{\,\bop\,}

\newcommand{\dop}{\de}
\newcommand{\dopx}{\,\dop^\vecx}

\newcommand{\dopxsin}{\,\dop^\vecx_0}
\newcommand{\dopxsie}{\,\dop^\vecx_1}
\newcommand{\dopxsii}{\,\dop^\vecx_i}
\newcommand{\dopxsiip}{\,\dop^\vecx_{i+1}}

\newcommand{\dopxsik}{\,\dop^\vecx_k}
\newcommand{\dopxsikp}{\,\dop^\vecx_{k+1}}

\newcommand{\dopy}{\,\dop^\vecy}

\newcommand{\lx}{\mathrm{S}^\vecx}

\newcommand{\ly}{\mathrm{S}^\vecy}
\newcommand{\sz}{\mathrm{sz}}
\newcommand{\lhx}{\mathrm{sz}^{\vecx}}

\newcommand{\txij}{\mathrm{T}^\vecx_{i,j}}

\newcommand{\txipj}{\mathrm{T}^\vecx_{i+1,j}}

\newcommand{\txnn}{\mathrm{T}^\vecx_{0,0}}

\newcommand{\txi}{\mathrm{T}^\vecx_{i}}
\newcommand{\txip}{\mathrm{T}^\vecx_{i+1}}

\newcommand{\subxij}{\mathrm{Sub}^\vecx_{i,j}}

\newcommand{\subxii}{\mathrm{Sub}^\vecx_{i,i}}

\newcommand{\subyij}{\mathrm{Sub}^\vecy_{i,j}}
\newcommand{\subyipj}{\mathrm{Sub}^\vecy_{i+1,j}}
\newcommand{\subynj}{\mathrm{Sub}^\vecy_{0,j}}
\newcommand{\subyej}{\mathrm{Sub}^\vecy_{1,j}}

\newcommand{\subykj}{\mathrm{Sub}^\vecy_{k,j}}

\newcommand{\subxipj}{\mathrm{Sub}^\vecx_{i+1,j}}

\newcommand{\subxne}{\mathrm{Sub}^\vecx_{0,1}}
\newcommand{\subxi}{\mathrm{Sub}^\vecx_{i}}
\newcommand{\subxip}{\mathrm{Sub}^\vecx_{i+1}}

\newcommand{\hij}{\{h\}_{i,j}}

\newcommand{\klamp}[1]{[\hspace{-1.5pt}[#1]\hspace{-1.5pt}]}
\newcommand{\klampn}[1]{[\hspace{-1.5pt}[#1]\hspace{-1.5pt}]_0}
\newcommand{\klampe}[1]{[\hspace{-1.5pt}[#1]\hspace{-1.5pt}]_1}

\newcommand{\val}{\mathrm{val}}

\begin{document}

\title[Derivation Lengths Classification of G\"odel's T]{Derivation Lengths Classification of G\"odel's T Extending Howard's Assignment\rsuper*}

\author[G.~Wilken]{Gunnar Wilken\rsuper a}
\address{{\lsuper a}Structural Cellular Biology Unit\\Okinawa Institute of Science and Technology\\
1919-1 Tancha, Onna-son, 904-0412 Okinawa, Japan}
\email{wilken@oist.jp}
\thanks{{\lsuper a}This work was partially supported by the Mathematical Biology Unit at OIST}

\author[A.~Weiermann]{Andreas Weiermann\rsuper b}
\address{{\lsuper b}Department of Mathematics\\ Ghent University\\ Building S22, 
Krijgslaan 281, B 9000 Gent, Belgium}
\email{weierman@cage.ugent.be}

\titlecomment{{\lsuper*}Complete rewrite of the article by Wilken and Weiermann in the Springer LNCS 5608 proceedings 
volume of TLCA 2009. Corresponding author is Wilken.}

\begin{abstract}
Let T be G\"odel's system of primitive recursive functionals
of finite type in the lambda formulation.
We define by constructive means using recursion on nested multisets
a multivalued function I from the set of terms of T into
the set of natural numbers such that if a term A reduces
to a term B and if a natural number I(A) is assigned
to A then a natural number I(B) can be assigned to B such
that I(A) is greater than I(B). The construction of I is based on Howard's
1970 ordinal assignment for T and Weiermann's 1998 treatment
of T in the combinatory logic version. As a corollary
we obtain an optimal derivation lengths classification
for the lambda formulation
of T and its fragments. Compared with Weiermann's 1998 exposition
this article yields solutions to several non-trivial problems
arising from dealing with lambda terms instead of combinatory logic
terms. 
It is expected that the methods developed here can
be applied to other higher order rewrite systems
resulting in new powerful termination orderings
since T is a paradigm for such systems.
\end{abstract}

\maketitle

\section*{Introduction}
This article is part of a general program
of investigations on subrecursive complexity
classes via derivation lengths classifications
of term rewriting systems. Quite often,
an equationally defined subrecursive complexity
class ${{\mathcal C}}$ of num\-ber-theoretic functions
can be defined in terms of a corresponding rewrite
system $R_{{\mathcal C}}$ which computes the functions
from ${{\mathcal C}}$. Appropriate bounds on the $R_{{\mathcal C}}$-derivation
lengths then yield intrinsic information on the computational
complexity of ${{\mathcal C}}$.
Successful examples of this program have been documented,
for example, in \cite{BW98,CW97}.
Having such applications in mind, it seems desirable
to have a large variety of powerful methods
for establishing bounds on derivation lengths
in general.

A common and very convenient tool for proving termination
of a reduction system consists
in defining an interpretation function $I$ from
the set of terms in question into the set of natural
numbers such that if a term $A$ rewrites to a term
$B$ then $I(A)>I(B)$.
A rewriting sequence of terms
$A_1\to\ldots\to A_n$
then yields a strictly descending chain of natural
numbers $I(A_1)>\ldots>I(A_n)$. The number $I(A_1)$ is thus an
upper bound for $n$ and hence the assignment function $I$ provides a termination
proof plus a non-trivial upper bound
on resulting lengths of longest possible reductions.
In this paper we apply a generalization
of this method -- the non-unique assignment technique --
to $\GTlam$, the $\lambda$-formulation of G\"odel's $\GT$,
which is the prototype for a higher order rewrite
system. For $\GTCL$, the combinatory logic formulation
of $\GT$, a corresponding interpretation has already been
constructed in \cite{W98}.
In this article we solve the technically more involved
problem of classifying the derivation lengths
for $\GTlam$ via a multivalued interpretation
function. The extra complications when compared
with the treatment in \cite{W98} are due to
the need for a variable concept underlying the assignment technique and to the 
simultaneous treatment of recursion, $\beta$-conversion,
and the $\xi$-rule. 

For a recent and extensive exposition of the history of termination proofs for
G\"odel's $\GT$ we refer the reader to Section 8.2 in \cite{CH}. In fact, \cite{CH} covers
the history of $\lam$-calculus and combinatory logic in general.
Unlike the present paper, the majority of termination
proofs for G\"odel's $\GT$ mentioned in \cite{CH} does not yield non-trivial upper bounds
on the lengths of reductions.

An alternative approach for proving termination of G\"odel's $\GT$ which yields
non-trivial upper bounds on the lengths of reductions was suggested in \cite{BW00}. There the lengths
of derivations were classified by proof-theoretic investigations
on head-reduction trees.

The current approach is more direct, and as a possible benefit for future work
we expect the extraction of powerful (syntactic) termination
orderings for higher order reduction systems which generalize the recursive path ordering.
We conjecture that such an
ordering for G\"odel's $\GT$ will have the Bachmann-Howard ordinal
as order type.

\subsection*{Preliminary Remarks}
Frequent mention of Howard's work \cite{H70} does not mean that we presuppose its knowledge. 
In fact, we have intended this paper to become as much self-contained 
as possible. Nevertheless, we have adopted much of the notation used in \cite{H70}
and \cite{W98}. Knowledge of those works together with \cite{BCW94,W96,W97} is certainly
useful in order to understand this article in greater depth, but it is not required.

Section \ref{GTsec} introduces G\"odel's $\GT$ in the typed $\lam$-calculus version together with
several well-known notions for its analysis. In addition, Subsection \ref{extsubsec}
will prove useful for the particular purposes of this paper, and Subsection 
\ref{assignsubsec} gives a heuristic explanation and motivation of the method of non-unique assignment. We believe that this facilitates the understanding of the method considerably.
Nevertheless, the results of Subsection \ref{assignsubsec} are not needed later on, hence
Subsections \ref{basiccons} and \ref{precisemotiv} can be skipped at first reading.
Subsection \ref{linksubsec} will clarify how \ref{assignsubsec} relates to our argumentation
in the central section of this paper, Section \ref{assignsec}. At that stage we expect the benefits from Subsections
\ref{assignsubsec} and \ref{linksubsec} to become fully convincing.
The ordinal theoretic means applied in Section \ref{assignsec} are introduced in Section 
\ref{ordinalsec}. Section \ref{classificationsec} provides a thorough analysis of the assignments given in Section \ref{assignsec},
showing that we obtain an exact classification of derivation lengths of $\GT$ and its fragments. 

The paper is organized in a way that allows for linear reading. However, detailed 
technical proofs required in our argumentation are given in the appendix in order 
to increase readability for an audience less familiar with ordinal theory.  

\section{Typed \boldmath$\lam$\unboldmath-Calculus with Recursion}\label{GTsec}

We give a short description of typed $\lam$-calculus extended by
recursors and case distinction functionals for each type. We will call this
variant {\it $\lbr$-calculus}.
The theory $\GTlam$ is G\"odel's $\GT$ based on $\lbr$-calculus.
We are going to slightly deviate from the notation of \cite{H70}.

\subsection{Types and Levels}\label{typelevelsubsec}

The set of {\it finite types} is defined inductively: it contains the type $0$ and the type $(\si\tau)$ whenever
$\si\tau$ are finite types. 
Type $0$ is intended to consist of the natural numbers whereas
$(\si\tau)$ for given finite types $\si$ and $\tau$ denotes
the type of functions $f : \si \to \tau$. 
We will denote finite types by the 
letters $\rho,\si,\tau,$ etc. In cases where ambiguity is unlikely, instead of $(\si\tau)$
we will simply write $\si\tau$, and we identify $\rho\si\tau$ with $\rho(\si\tau)$.
As in \cite{H70} the {\it level} of a finite type is defined recursively by
\[\lv(0) := 0,\quad\lv (\si\tau) := \max\{\lv(\si) + 1, \lv(\tau)\}.\]

\subsection{Terms, Subterms, Parse Trees, Substitution, $\alpha$-Equivalence}\label{termssubsec}

Let $\Var$ be a countably infinite set of variables which we denote by $X, Y, Z$, etc.
Following the Church-style convention, we choose a type for each of those variables
in such a way that $\Var$ contains infinitely many variables of each type. This choice
is fixed throughout the paper, and we will sometimes indicate the type, say $\si$, of a variable $X$ by the notation $\xsi$.
 
The set $\T$ of {\it typed terms} is defined inductively. It contains
\begin{iteMize}{$\bullet$}
\item all variables in $\Var$,
\item a constant $\N$ of type $0$,  
\item a constant $\Suc$ of type $00$,   
\item case distinction functionals\footnote{Case distinction functionals are needed to the full extent in the functional interpretation of the fragments $\mathrm{I}\Sigma_{n+1}$ of Peano Arithmetic in the fragments $\GT_n$ of $\GT$, 
cf.\ \cite{P72}.}  $\Dt$ of type $0\tau\tau\tau$, 
\item primitive recursion functionals $\Rt$ of type $0(0\tau\tau)\tau\tau$ for each $\tau$, 
\end{iteMize}
and is closed under
\begin{iteMize}{$\bullet$}
\item application, that is, $(AB)$ is a term of type $\tau$ whenever $A$ is a term of type $\si\tau$ and $B$ is a term of type $\si$ and 
\item abstraction, that is, $\lam X.B$ is a term of type $\rho\si$
whenever $X$ is a variable of type $\rho$, $B$ is a term of type $\si$.
\end{iteMize}
We will suppress parentheses whenever ambiguity is unlikely to arise, e.g.\ we write
$AB$ instead of $(AB)$. We further identify $ABC$ with $(AB)C$ for any terms $A, B, C$ of suitable types.
If a term $A$ is of type $\tau$ we communicate this sometimes
by writing $A^\tau$. Conversely, instead of writing $\Dt, \Rt$ we sometimes simply write $\D, \R$, respectively. We continue  to denote terms of $\T$ by Roman capital letters with the exception of recursion arguments in terms of the form $\R t$, where we sometimes use lower case Roman letters $s,t,$ etc.

The set $\BV(A)$ of {\it bound variables} of a term $A$ consists of all variables $X$ for which $\lam X$ occurs in 
$A$, whereas the set $\FV(A)$ of {\it free variables} of $A$ consists of all variables $X$ occurring in $A$
outside the scope of $\lam X$. 

The set of {\it subterms} of a given term $A$ is defined as usual by induction on the buildup of $A$, including $A$ itself.
The {\it direct  subterms} of terms of a form $AB$ are $A$ and $B$, the only direct  subterm of 
$\lam X.A$ is $A$, all other terms do not have any direct  subterm. 

The {\it parse tree} of a term $A$ is the labeled tree whose root is labeled with $A$, 
and whose immediate subtrees are the parse trees of the immediate subterms of $A$ (if any).
The set of all labels of nodes of the parse tree of a term $A$ is therefore equal to the set of subterms of $A$. The 
parse tree, however, distinguishes between possibly various occurrences of the same subterm of $A$ and provides their contexts within $A$.
The set of nodes of the parse tree of a term $A$ can be identified with the {\it addresses}
of $A$, which are finite strings over $\{1,2\}$ according to the following definition,
where $\epsilon$ denotes the empty string.
The {\it subterm of $A$ at address $a$}, written $A_{\mid a}$, is defined inductively by
\begin{iteMize}{$\bullet$}
\item $A$ if $a=\epsilon$,
\item $B_{\mid b}$ if $A = \lam X.B$ and $a=1b$,
\item $B_{\mid b}$ if $A = BC$ and $a=1b$,
\item $C_{\mid c}$ if $A = BC$ and $a=2c$.
\end{iteMize} 

{\it Substitution} $A\{X:=B\}$ is defined as usual by induction on the buildup of $A$, replacing any free occurrence of the variable $X$ in $A$ by the term $B$. The variable condition $\BV(\lam X.A) \cap \FV(B) = \emptyset$ for 
$\be$-conversion (see below) avoids variable capture. In other words, it makes sure 
that none of the free variables in
$B$ becomes bound within $A\{X:=B\}$. Moreover, the abstraction variable $X$
does not occur in $A\{X:=B\}$. 

A term $A$ is called {\it well-named} iff for each subterm $B$ of $A$ we have
$\FV(B)\cap\BV(B)=\emptyset$, and furthermore, for each variable $X$, the binder 
$\lam X$ occurs at most once in $B$.

Let $A$ be a term with subterm $\lam X.B$ at address $a$.
If $Y$ is a variable of the same type as $X$ such that $Y\not\in\FV(B)$, 
then we say that $A$ {\it $\alpha$-converts} to the term resulting from $A$ by the replacement of $A_{\mid a}$ by $\lam Y.(B\{X:=Y\})$.
If a term $C$ is obtained from $A$ via a finite number of $\alpha$-conversions, then
we call $A$ and $C$ {\it $\alpha$-equivalent}, $A=_\alpha C$.
Notice that every term is $\alpha$-equivalent to some well-named term. 

\subsection{Equivalence and  \boldmath$\lbr$-Reduction\unboldmath}

The {\it equivalence relation} is the reflexive, symmetric and transitive
closure of $\alpha$-equivalence and the {\it one-step reduction} $\rhd$, defined 
inductively as least binary relation on terms such that
\[\begin{array}{ll@{\quad\quad}l@{\;\;}l}
  (D_0) & \D\N AB \rhd  A & (D_{\Suc}) & \D(\Suc t)AB \rhd B\\[1mm]
  (R_0) & \R\N AB \rhd B & (R_{\Suc}) & \R(\Suc t)AB \rhd At(\R tAB)\\[1mm]
  (App_r) & A \rhd B \imp AC \rhd BC & (App_l) & B \rhd C \imp AB \rhd AC\\[1mm]
  (\be) & \multicolumn{3}{l}{(\lam X.A)B \rhd A\{X:=B\}
        \mbox{ where } \BV(\lam X.A) \cap \FV(B) = \emptyset}\\[1mm]
  (\xi) & \multicolumn{3}{l}{A \rhd B \imp \lam X.A \rhd \lam X.B}
\end{array}\]

Clearly, the variable condition for $\beta$-conversion is satisfied if the term
$(\lam X.A)B$ is well-named. Note that $\rhd$ does not preserve well-namedness. 
However, as mentioned above, well-namedness can always be restored: if $A\rhd B^\prime$
then there exists a well-named term $B$ such that $A\rhd B^\prime=_\alpha B$.
On well-named terms we define $A\rhdal B$ iff there exists (the unique) $B^\prime$ such
that $A\rhd B^\prime=_\alpha B$.

The calculus $\lbr$ enjoys the Church-Rosser property (confluence) and strong normalization, and every functional of type $0$ reduces to a numeral (cf.\ \cite{HinSel86}).

Note that extending $\lbr$ by $\eta$-contractions would cause loss of the Church-Rosser property, consider for example the term $\lam X.\R0AX$.

\subsection{Extension to \boldmath$\Tpr$ and $\lbrpr$\unboldmath}\label{extsubsec} 

In order to prepare a separate treatment of $\R$-reductions and $\be$-reductions, which otherwise would be
incompatible within the framework of our intended non-unique assignment, we introduce  
the following definitional extension of $\lbr$-calculus. We extend the clauses in \ref{termssubsec}
defining the terms of $\T$ so that 
\begin{iteMize}{$\bullet$}
\item $\R^t$ is a term of type $(0\tau\tau)\tau\tau$ for any $\Rt$ and any term $t$ of type $0$.
\end{iteMize} 
We call the extended set of terms $\Tpr$.
The terms $\R^t$ are special forms of application terms and thus allow for a clean way to give two 
different assignments to terms resulting from the application of some $t$ to a recursor $\R$. 
The effect will be that assignments to $\R t$ allow for a treatment of $\be$-conversion while assignments to
$\R^t$ allow for a treatment of $\R$-reduction. 
This smoothly fits with our treatment of the $\xi$-rule by the method of 
non-unique assignment according to \cite{H70}, which takes advantage from the additional information given by the reduction history of terms, see Section \ref{assignsubsec}.

The subterms of $\R^t$ are $\R^t$ itself, $\R$, and the subterms of $t$. The direct  subterms are $\R$ and $t$. Parse trees for the extension are defined accordingly, $(\R^t)_{\mid a}$
is defined by $(\R t)_{\mid a}$.
Substitution for the new terms is defined by \[\R^t\{X:=A\}:\equiv\R^{t\{X:=A\}},\]
and $\alpha$-equivalence is expanded to all terms of $\Tpr$.
The one-step reduction relation is then modified to include additional rules $(R)$ and $(App_R)$, and $(R_0)$ and 
$(R_{\Suc})$ are replaced by $(R^0)$ and $(R^{\Suc})$, respectively, as follows: 
\[\begin{array}{ll@{\quad\quad}l@{\;\;}l}
(R) & \R t \rhd \R^t & (App_R) & s \rhd t \imp \R^s \rhd \R^t\\[1mm]
(R^0) & \R^\N AB \rhd B & (R^{\Suc}) & \R^{\Suc t}AB \rhd At(\R^t AB).
\end{array}\]
We denote the modified calculus for the terms of $\Tpr$ by $\lbrpr$.
The properties and definitions regarding $\lbr$ mentioned in the previous subsection are easily seen to carry over to $\lbrpr$.

We define the {\it length} of a term $A$, $\lh(A)$, as follows:
\begin{iteMize}{$\bullet$}
\item $\lh(A):=1$ if $A$ is a variable or constant, 
\item $\lh(BC):=\lh(B)+\lh(C)$, 
\item $\lh(\R^t):=1+\lh(t)$, and 
\item $\lh(\lam X.G):=\lh(G)+1$.
\end{iteMize}

It is a trivial observation that by identifying any $\R^t$ with $\R t$ and omitting all $(R)$-reductions we recover the original $\lbr$-calculus where $(App_R)$-reductions turn into $(App_l)$-reductions and $(R^0)$, $(R^\Suc)$ turn into $(R_0)$, $(R_\Suc)$, respectively. 
Clearly, reduction sequences can only become shorter in this process. 

On the other hand, given any reduction sequence in $\lbr$, we obtain a corresponding reduction
sequence in $\lbrpr$ (of at most double length) by straightforward insertion of $(R)$-reductions as needed, 
from which we recover the original sequence by the process described above.   

Another trivial observation is that given a reduction sequence in $\lbrpr$, starting from a term $A\in\Tpr$ we can straightforwardly find a term $A^\prime\in\T$ and a reduction sequence which transforms $A^\prime$ into $A$ in at most $\lh (A)$-many steps, only using 
$(R)$-reduction.  
It is therefore sufficient to perform a derivation lengths classification for the class of reduction sequences in
$\lbrpr$ which start from $\T$-terms, as the resulting bounds will turn out to be sharp for $\lbr$. 

\subsection{Reduction Trees} 

The rules $(App_l),(App_r),(\xi)$, and $(App_R)$ imply that the one-step reduction $\rhd$ applied to some term $A$ consists of the reduction of one redex of $A$. A redex of $A$ is the occurrence of a subterm of $A$ (corresponding to a unique node in the parse tree of $A$) that matches any left-hand side of the reduction clauses for $\D$-, $\R$-, or $\be$-reduction. Inside $A$ the redex is then replaced by the (properly instantiated) right-hand side of the corresponding clause.
We call the redex chosen for a particular one-step reduction $\rhd$ the {\it working redex}. The {\it reduction tree} of a (well-named) term $A$ is then given by the exhaustive
application of $\rhdal$ in order to pass from a parent to a child node, starting from $A$. We do not need to be any more specific about the arrangement of 
nodes in reduction trees.    

\subsection{Assignment of Ordinals to Terms of \boldmath$\Tpr$\unboldmath}\label{assignsubsec} 
\subsubsection{Overview}
The aim of the present paper is to give exact bounds on the heights of reduction trees in the usual sense of derivation lengths classification for term rewriting systems. This is achieved by the assignment of strictly decreasing natural numbers to the terms of reduction sequences. The natural numbers assigned to terms are computed along vectors 
of ordinal terms (from upper down to lower components) which in turn are built over variable vectors that correspond to typed variables. 

The assignment is not unique for terms because it is dependent on the respective reduction history. The main reason for this dependency on reduction histories is the same as already
encountered in \cite{H70}. The treatment of $\beta$-reduction involves an operator on
ordinal vectors which is not monotone with respect to the $\xi$-rule. The other reason is 
the incompatibility of our treatments of $\R$- and $\beta$-reduction, as already mentioned in
Subsection \ref{extsubsec}. The solution outlined there is based on the fact that when
considering $\lbrpr$-reduction sequences which start from terms in $\T$,
we can trace back abstraction subterms in the 
reduction history, obtaining {\it corresponding} subterms, cf.\ Definition 
\ref{correspondencedefi}, in which the respective abstraction variable does not occur in subterms of the form $\R^t$. 
This observation is crucial for our simultaneous treatment of $\beta$-reductions, the $\xi$-rule, and {\it arbitrary} 
$\R$-reductions, which were excluded in \cite{H70}.
The assignment to such corresponding terms earlier in the reduction history is then the
key to the handling of $\beta$-reductions that may occur much later in the reduction sequence.

Now, given the particular reduction history of a term $A$ occurring in a fixed reduction sequence, the assignment is determined uniquely
(we will introduce the crucial notion of {\it assignment derivation} in our formal argumentation) and built up from the 
assignments to the nodes of the parse tree of $A$ and terms occurring earlier in the reduction history of $A$. 
Additionally, the assembly of new assignments along the reduction
sequence involves (iterated) substitutions of variable vectors by already defined assignments, generating terms which do {\it not} occur
as subterms of terms in the reduction history of $A$.  
For this reason, besides the obvious reason in the treatment of $\beta$-reduction, the assignment method 
has to be designed so as to naturally commute with substitution, as was done already in \cite{H70}. Another essential property
of our assignment method is its invariance under $\alpha$-equivalence, as in \cite{H70}. 
This will enable us to treat $\rhdal$-reductions in the same way as $\rhd$-reductions.

Our construction of assignments to terms will start from unique assignments to all 
variables and constants, using Howard's operator $\bop$ in its refined form of
\cite{W98} to compute an assignment to a term $BC$ from assignments to $B$ and $C$, a specific treatment of terms $\R^t$ as used 
in \cite{W98}, where it was used for terms of the form $\R t$, and a refinement of Howard's operator $\dop$,
cf.\ \cite{H70}, to be applied in the treatment of abstraction terms, which causes the non-uniqueness of the
assignment method.

\subsubsection{Basic Considerations}\label{basiccons}
Given a reduction $A\rhd B$ we begin with describing how the parse tree for $B$ is obtained from the parse tree for $A$ in a uniform way.
This will be crucial for the construction of our assignment.
Focusing on the working redex of the reduction $A\rhd B$ we consider the path $\pathP$ from the root, labeled 
with $A$, to the working redex at node $r$, labeled with $F$.
Clearly, $F$ is a subterm of any term labeling a node of $\pathP$. Assume the working redex is reduced
(via $\D$-, $\R$-, or $\be$-reduction) to the term $G$ at node $s$ in the parse tree of $B$.
The subtree with root $r$ of the parse tree of $A$, that is the parse tree for $F$, is replaced with the parse tree for $G$, and along the path $\pathP$ the labels are modified by the replacement of the working redex $F$ by the 
reduct $G$. If the meaning is clear from the context we will sometimes denote such a replacement by $H[s/r]$ 
where $H$ is a label of a node on $\pathP$.
All remaining nodes of the parse tree of $A$ are preserved in the parse tree of $B$ with
the same labeling term. In other words, the tree structure is modified by the replacement of the parse tree of $F$ at node $r$ by the parse tree of $G$ with the corresponding labeling, while the modification of labels additionally involves the labels along $\pathP$ in form of the replacement $[s/r]$ of the working redex $F$ by $G$. 

In the case of $\D$- and $\R$-reductions the transformation of the parse tree of $F$ to the parse tree of $G$ is clear, and we can uniquely identify subtrees
of redex and reduct, including their labels. For example in the case where 
$F$ is a term $\R^{\Suc t}CD$ and 
$G$ is the term $Ct(\R^tCD)$, we can trace the parse tree of 
$F$ until we reach the parse trees of $t$, $C$, and $D$, and do the same with the parse tree of $G$, identifying the
corresponding subtrees. We are going to use this identification of subtrees
in our assignment in order to carry over already defined assignments. 

Now consider the case where the working redex is a $\be$-redex, say $F$ is $(\lam X.C)D$ and 
$G$ is $C\{X:=D\}$.
The immediate subtrees of the parse tree of $F$ are then the parse trees of $\lam X.C$ and $D$, the immediate subtree of the former being the parse tree $\treeS$ of $C$.
Consider the tree $\treeS\{X:=D\}$ which is obtained as follows. The tree structure is obtained from $\treeS$ by replacing every leaf with label $X$ by the parse
tree of $D$ with the corresponding labels. The remaining labels are modified by the substitution $\{X:=D\}$. The subtrees substituted for the leaves with label
$X$ can be identified with the immediate subtree of $F$ which is the parse tree of $D$.

Having discussed the transition from the parse tree of $A$ to the parse tree of $B$, where
$A\rhd B$, let us now assume that 
$Y\in\FV(B)$ and that $C$ is a term of the same type as $Y$ such that $\BV(B)\cap\FV(C)=\emptyset$.
Modulo $\al$-congruence we may assume that also $\BV(A)\cap\FV(C)=\emptyset$.
We then clearly have $Y\in\FV(A)$, and $A\{Y:=C\}\rhd B\{Y:=C\}$. 
The parse trees of $A\{Y:=C\}$ and
$B\{Y:=C\}$ are obtained from those for $A$ and $B$, respectively, by replacing every leaf with label $Y$
by the parse tree of $C$ and by the modification of all remaining labels by the substitution $\{Y:=C\}$. 

\subsubsection{Precise Motivation}\label{precisemotiv}
Bearing the above preparation in mind we proceed with a precise explanation of the method of non-unique assignment 
that was used in Section 4 of \cite{H70} in order to handle the unrestricted $\xi$-rule. 
As mentioned above, the approach of non-unique assignment had to be refined so as to manage arbitrary $\R$-reductions.
 
\begin{defi}\label{correspondencedefi}
Let $A,B$ be terms such that $A\rhd B$ with working redex and reduct at address $w$,
respectively. Let $b$ be an address in $B$ such that $b=q1$ and $B_{\mid q}$ is an abstraction. We define a unique address $a$ and say that {\it $a$ corresponds to $b$} 
w.r.t.\ the pair $(A,B)$ as follows.
\begin{enumerate}[(1)]
\item If $b$ and $w$ are incomparable, then we have $A_{\mid b}=B_{\mid b}$ and set $a:=b$.
\item If $b$ is a prefix of $w$ (written as $b\subseteq w$), then we have 
$A_{\mid b}\rhd B_{\mid b}$ and set $a:=b$.
\item If $w$ is a proper prefix of $b$, that is, $w\subsetneq b$. 
\begin{enumerate}[(3.1)]
\item[(3.1)] $w\not=\epsilon$. Then let $b^\prime$ be such that $b=wb^\prime$, 
let $a^\prime$ be such that $a^\prime$ corresponds to $b^\prime$ w.r.t.\ 
$(A_{\mid w},B_{\mid w})$, and set $a:=wa^\prime$.
\item[(3.2)] $w=\epsilon$. Then we distinguish between the following cases:
\begin{enumerate}[(a)]
\item $A=(\lam X.C)D$, $B=C\{X:=D\}$. 
\begin{enumerate}[(i)]
\item If $b$ is of a form $cd$ such that $C_{\mid c}=X$ then $a:=2d$, 
\item otherwise $a:=11b$.
\end{enumerate}
\item $A=\R t$, $B=\R^t$. Then set $a:=b$.
\item $A=\R^0CB$. Set $a:=2b$.
\item $A=\R^{\Suc t}CD$, $B=Ct(\R^tCD)$. 
\begin{enumerate}\renewcommand{\labelenumiv}{(\roman{enumiv})}
\item If $b$ is of a form $22d$, then $a:=2d$.
\item If $b$ is of a form either $11c$ or $212c$, then $a:=12c$.
\item If $b$ is of a form either $12e$ or $2112e$, then $a:=1122e$.
\end{enumerate}
\item $A=\D0BC$. Set $a:=12b$.
\item $A=\D(\Suc t)CB$. Set $a:=2b$.
\end{enumerate}
\end{enumerate}
\end{enumerate}

\noindent In case (2) we call the reduction $A_{\mid b}\rhd B_{\mid b}$ the {\it associated reduction} w.r.t.\ $(A,B)$ and $b$, otherwise an associated reduction is not defined.

If the working redex is a $\be$-redex, say $A_{\mid w}=(\lam X.C)D$, and $w\subsetneq b$, $b$ not of a form
$b=wcd$ with $C_{\mid c}=X$, then we call the substitution $\{X:=D\}$ the 
{\it associated substitution} w.r.t.\ 
$(A,B)$ and $b$, otherwise an associated substitution is not defined.
 
For well-named terms $A, B$ such that $A\rhdal B$ and addresses $a,b$ in $A, B$, respectively, let $B^\prime$ be the unique term such that $A\rhd B^\prime=_\al B$. 
Then we say that $a$ corresponds to $b$ w.r.t.\ the pair $(A,B)$ iff
$a$ corresponds to $b$ w.r.t.\ the pair $(A,B^\prime)$. The associated reduction 
(respectively, associated substitution) w.r.t.\ 
$(A,B)$ and $b$ is defined iff it is defined for $(A,B^\prime)$ and $b$, in which case they
are the same.
\Fin
\end{defi}

Note that for any $A,B,a,b$ as in the above Definition, $a$ is of the form
$p1$, and $A_{\mid p}$ is an abstraction. Notice further that exactly one
of the following holds: 
\begin{enumerate}[(I)]
\item The associated reduction is defined.
\item The associated substitution is defined.
\item Neither the associated reduction nor the associated substitution is defined. 
\end{enumerate}
Notice that if $A\rhd B$, in case (I) we have $A_{\mid a}\rhd B_{\mid b}$,
in case (II) we have $A_{\mid a}\{X:=D\}=B_{\mid b}$ where $\{X:=D\}$ is the associated substitution, and
in case (III) we have $A_{\mid a}=B_{\mid b}$.

However, if we have $A\rhdal B$, say $A\rhd B^\prime=_\al B$,
in general the terms $B_{\mid b}$ and $B^\prime_{\mid b}$ do not only differ due to
the renaming of bound variables but also due to renaming of free variables as they
might be subterms of abstractions that have undergone $\alpha$-conversion.
In particular, it does not make sense to keep track of the abstraction variable of 
$B_{\mid q}$.
The following lemma addresses this technical issue.
 
\begin{lem}\label{alphalem}
Let $A,B,C$ be terms such that $A\rhd B=_\alpha C$ and suppose that $A,C$ are well-named
and that the abstraction variables used in order to $\al$-convert $B$ to $C$ do neither
occur free nor bound in $A$ or $B$.
Let $b=q1$ be an address such that $C_{\mid q}$ is an abstraction
term, say $\lam X.H$, and let $a=p1$ be the corresponding address in $A$.
Then there exists a well-named term $A^*$ such that
\begin{enumerate}[\em(1)] 
\item $A^*=_\al A$,
\item $A^*\rhd B^* =_\al C$,
\item\label{crucialprop} $B^*_{\mid q}=\lam X.(B^*_{\mid b})$ where $B^*_{\mid b}=_\al H$, and 
\item exactly one of the following holds:
\begin{enumerate}[\em(a)]
\item $A^*_{\mid a}\rhd H$,
\item $A^*_{\mid a}\{Y:=D\}=_\al H$ 
where $\{Y:=D\}$ is the assoc.\ substitution w.r.t.\ $(A^*,C)$,
\item $A^*_{\mid a}=_\al H$. 
\end{enumerate}
\end{enumerate}
\end{lem}
\proof Notice first of all that (2) follows from (1) and $B^*$ is determined by (2). 
Suppose $\{v_1,\ldots,v_m\}$ is the set of addresses in $B$ at which
$\al$-conversion has to be performed in order to obtain $C$ (the order of those
$\al$-conversions is of course irrelevant), with corresponding new variables
$V_1,\ldots,V_m$. Let $w$ be the address of the working redex in $A$ and hence 
also the address of the reduct in $B$. In the case $m=0$ we choose $A^*:=A$ and are
done. 

Now suppose that $m>0$ and consider $v\in\{v_1,\ldots,v_m\}$ with 
corresponding new abstraction variable $V$. We investigate whether an $\al$-conversion
in $A$ becomes necessary in order to satisfy the lemma. The collection of all 
$\al$-conversions that have to be performed in $A$ will then determine the term
$A^*$.

If $v$ is not a prefix of $q$, then the $\al$-conversion at $v$ in $B$ does not
cause any change in $C_{\mid q}$ and hence does not require any additional 
$\al$-conversion in $A$.

Suppose now that $v$ is a prefix of $q$, that is, $v\subseteq q$. We then consider
3 cases regarding the addresses $v$ and $w$.

{\bf Case 1:} $v$ and $w$ are incomparable.
Then we have $p=q$, $v1$ corresponds to $v1$ w.r.t.\ $(A,B)$, and we draw the 
$\al$-conversion at $v$ in $B$ back to an $\al$-conversion at $v$ in $A$ with the 
same new variable $V$. 

{\bf Case 2:} $v\subsetneq w$. Then $A_{\mid v}$ is of a form $\lam Z.G$
and $A_{\mid v}\rhd B_{\mid v}=\lam Z.G^\prime$. 
Hence $C_{\mid v}=\lam V.G^\prime\{Z:=V\}$. Again 
$v1$ corresponds to $v1$ w.r.t.\ $(A,B)$, and we draw the 
$\al$-conversion at $v$ in $B$ back to an $\al$-conversion at $v$ in $A$ with the 
same new variable $V$. The subterms of $A_{\mid v}$ are now subject to the 
variable substitution $\{Z:=V\}$, which includes the working redex and the term
$A_{\mid p}$.
 
{\bf Case 3:} $w\subseteq v$. Then we have $w\subseteq v\subseteq q$. 
Let $u$ be the address in $A$ such that $u1$ corresponds to $v1$ w.r.t.\ $(A,B)$.
We have $v1\subseteq b$ and $w\subsetneq u$ (abstractions cannot be working
redices). We perform the $\al$-conversion at $u$ in $A$ switching to the abstraction 
variable $V$. The reduction of $A$ to $B$ might generate further copies of $A_{\mid u}$
in $B$ whose addresses in $B$, however, cannot be prefixes of $q$. In $B^*$ we will have
corresponding copies, differing from those in $B$ by the abstraction variable $V$.
Hence $\al$-conversions
of those modified copies in $B^*$ to fit the corresponding abstractions in $C$ are possible and
determined by $C$.  
\qed
 
\begin{defi}\label{tracedefi} 
Let $F_0\rhd\ldots\rhd F_n$ be a reduction sequence and let $p$ be the address 
identified with a node in the parse tree of $F_n$ that is 
labeled with an abstraction term $\lam X.H$. 
The {\it trace} of $p$ in $F_0\rhd\ldots\rhd F_n$ is the sequence $(b_0,\ldots,b_n)$
such that $b_n=p1$ and each $b_i$ corresponds to $b_{i+1}$ w.r.t.\ $(F_i,F_{i+1})$.
The {\it associated trace terms} $H_0,\ldots,H_n$ are the terms 
$F_{0\mid b_0},\ldots,F_{n\mid b_n}$,
thus $H_n=H$.
Let a partition $I_R,I_S,I_E$ of the index set $\{0,\ldots,n-1\}$ be defined as follows
\begin{iteMize}{$\bullet$}
\item $i\in I_R$ if the associated reduction w.r.t.\ $(F_i,F_{i+1})$ and $b_{i+1}$ is defined,
\item $i\in I_S$ if the associated substitution w.r.t.\ $(F_i,F_{i+1})$ and $b_{i+1}$ is defined, and
\item $i\in I_E$ otherwise,
\end{iteMize} 
according to the remark following Definition \ref{correspondencedefi}.
The index set $I_S$ therefore gives rise to the {\it associated partial substitution list}.

For a reduction sequence $F_0\rhdal\ldots\rhdal F_n$ we define the same notions, proceeding
as in Definition \ref{correspondencedefi}, and defining the trace terms $H_i$ by
$F_{i\mid b_i}$.
\Fin
\end{defi}

Given a reduction sequence $F_0\rhdal\ldots\rhdal F_n$ we have
unique terms $F^\prime_1,\ldots,F^\prime_n$ such that 
\[F_0\rhd F^\prime_1=_\al F_1\ldots F_{n-1}\rhd F^\prime_n=_\al F_n.\]
Assume that each conversion from $F^\prime_{i+1}$ to $F_{i+1}$ only uses new variables,
i.e.\ variables that have not been used earlier in the reduction sequence.
Clearly, this requirement on the reduction sequence can always be obtained via
$\al$-equivalence (possibly including renamings of bound variables in $F_n$).
Given an address $p$ identified with a node in the parse tree of $F_n$ that is 
labeled with an abstraction term $\lam X.H$, let $(b_0,\ldots,b_n)$ be the trace of $p$
and $I_R,I_S,I_E$ be the partition of the set of indices $\{0,\ldots,n-1\}$ according to
Definition \ref{tracedefi}.

Let $F^*_n:=F_n$. Iterated application of Lemma \ref{alphalem} starting from 
$F_{n-1}\rhd F^\prime_n=_\al F_n$ yields terms $F^*_{n-1},\ldots,F^*_0$ such that
\begin{enumerate}[(1)]
\item $F_i=_\al F^*_i$ for each $i$,
\item $F^*_0\rhdal\ldots\rhdal F^*_n$ with the same trace of $p$, the same partition 
$I_R,I_S,I_E$, and trace terms $H_i:=F^*_{i\mid b_i}$, and
\item for each $i<n$ we have: 
\begin{enumerate}[(a)]
\item $H_i\rhd H_{i+1}$ if $i\in I_R$,
\item $H_i\{Y_i:=D_i\}=_\al H_{i+1}$ if $i\in I_S$ where $\{Y_i:=D_i\}$ is the associated substitution w.r.t.\ $(F^*_i,F^*_{i+1})$, or
\item $H_i=_\al H_{i+1}$ if $i\in I_E$.
\end{enumerate}
\item The associated partial list of substitutions w.r.t.\ $F^*_0\rhdal\ldots\rhdal F^*_n$
and $p$ is {\it $X$-free}, that is, none of the terms $D_i$ which is defined
contains the variable $X$ and none of the defined variables $Y_i$ is identical with $X$.
\item Each $b_i$ is of a form $b^\prime_i 1$ and $F^*_{i\mid b^\prime_i}=\lam X.H_i$. 
\end{enumerate}

\begin{defi}
For a reduction sequence $F_0\rhdal\ldots\rhdal F_n$ with the above specified condition
of freshness of abstraction variables introduced by $\al$-conversions and node $p$ in $F_n$ as above, 
we call the sequence $F^*_0\rhdal\ldots\rhdal F^*_n$ defined above the {\it $p$-companion} of 
$F_0\rhdal\ldots\rhdal F_n$.
If $F_0\rhdal\ldots\rhdal F_n$ can be a $p$-companion of itself, then we call it {\it nice} w.r.t.\ $p$.
\Fin
\end{defi}

Notice that the crucial property for $p$-niceness of reduction sequences roots in
property \ref{crucialprop} of Lemma \ref{alphalem}: the invariance of the free variables 
which occur in the trace terms.

\begin{defi}\label{assocredseqdefi} 
Let $F_0\rhdal\ldots\rhdal F_n$, $n>0$, be a nice reduction sequence w.r.t.\   
a node $p$ in the parse tree of $F_n$ that is labeled with an abstraction term $\lam X.H$.
Let further $(b_0,\ldots,b_n)$ be the trace of $p$ in $F_0\rhdal\ldots\rhdal F_n$ 
with associated  trace terms $H_0,\ldots,H_n$ and 
associated partial list of substitutions $(\{Y_i:=D_i\})_{i\in I_S}$.

For each $i\in\{0,\ldots,n-1\}$ we define terms $\{H^j_i\}_{i\le j\le n}$ such that
\begin{iteMize}{$\bullet$}
\item $H^i_i:=H_i$,
\item $H^{j+1}_i=_\al C^j_i\{Y_j:=D_j\}$ if $j\in I_S$, where
\begin{iteMize}{$-$}
\item $C^j_i=_\al H^j_i$,
\item $H^{j+1}_i$ and $(\lam Y_j.C^j_i)D_j$ are well-named,
\end{iteMize}
using only fresh variables for $\al$-conversion,
\item $H^{j+1}_i$ otherwise.
\end{iteMize}

Let $(i_j)_{0\le j\le m}$ be the longest sequence such that
\begin{iteMize}{$\bullet$}
\item $i_0=0$,
\item $i_{j+1}$ is the least $i>i_j$ such that $H_{i-1}\rhd H_i$.
\end{iteMize}

We now let $G_m:=H_n$, define $G_j:=H^n_{i_j}$ for each $j\in\{0,\ldots,m-1\}$,
and call the sequence $G_0\rhdal\ldots\rhdal G_m$ the {\it associated reduction sequence} w.r.t.\   
$F_0\rhdal\ldots\rhdal F_n$ and $p$. 

In the case of the trivial reduction sequence $F_0$ with node $p$ in the parse tree of $F_0$ that is labeled with $\lam X.H$ we
define $G_0:=H$ to be the associated reduction sequence w.r.t.\ $F_0$ and $p$.
\Fin 
\end{defi}
Notice that in the above definition $H^n_{i_m}=_\al H_n$ which justifies our choice of $G_m$.
We have now carefully shown how, making
extensive use of $\al$-conversion, abstraction terms can be traced in reduction sequences. This will be essential in
the treatment of the $\xi$-rule. 
As mentioned before, our assignments to terms will be
invariant modulo $\al$-congruence.    

The inductive definition below introduces a binary relation $\wt$ originating from $t$ on p.457 of \cite{H70}, which assigns weights to terms of $\Tpr$ in a non-unique manner. For the purpose of the present section the assignment of weights serves as a useful illustration,
however, the use of weights will not be required later on.

\begin{defi}\label{wdefi}
Let $\wt$ be the minimal binary relation on $\Tpr\times\Nat^+$ which satisfies
\begin{iteMize}{$\bullet$}
\item $\wt(A,1)$ if $A$ is a variable or constant,
\item $\wt(BC,n)$ if $\wt(B,n_B)$, $\wt(C,n_C)$, and $n=n_B+n_C$,
\item $\wt(\R^t,n+1)$ if $\wt(t,n)$,
\item $\wt(\lam X.G,n)$ if there are terms $G_0\rhdal\ldots\rhdal G_m=G$ such that
$X$ does not occur in any subterm of $G_0$ of a form $\R^t$, $\wt(G_i,n_i)$ for $i=0,\ldots,m$,
and $n=1+n_0+\ldots+n_m$. 
\end{iteMize}
Every relation $\wt(F,n_F)$ for some $F\in\Tpr$ comes with a witnessing {\it derivation}, 
which is a tree whose root is labeled with $\wt(F,n_F)$, defined as follows:  
\begin{iteMize}{$\bullet$}
\item If $A$ is a variable or constant, then the tree consisting only of its root is the derivation of $\wt(A,1)$.
\item For derivations $\treeR$ of $\wt(B,n_B)$ and $\treeS$ of $\wt(C,n_C)$, the tree with direct subtrees $\treeR$ and $\treeS$ 
is a derivation of $\wt(BC,n_B+n_C)$.
\item For a derivation $\treeR$ of $\wt(t,n)$, the tree with direct subtree $\treeR$ is a derivation of $\wt(\R^t,n+1)$. 
\item For derivations $\treeR_0,\ldots,\treeR_m$ of $\wt(G_0,n_0),\ldots,\wt(G_m,n_m)$ with $G_0\rhdal\ldots\rhdal G_m=:G$ such that
$X$ does not occur in any subterm of $G_0$ of a form $\R^t$, the tree with direct subtrees
$\treeR_0,\ldots,\treeR_m$ is a derivation of $\wt(\lam X.G,1+n_0+\ldots+n_m)$. \Fin 
\end{iteMize}\smallskip
\end{defi} 

\noindent In Section \ref{assignsec}, instead of weights we will assign ordinal vectors to terms,
resulting in {\it assignment derivations}, see Definition \ref{assigndefi}. 
The notion of assignment derivation will be {\it more restrictive} than the notion of derivation here.
Reduction sequences $G_0\rhdal\ldots\rhdal G_m= G$ as mentioned above for the treatment of an abstraction $\lam X.G$
will additionally have to have a strictly decreasing sequence of assignments, which is crucial in the treatment of the $\xi$-rule.

We are going to show how weights can be assigned to terms along any reduction sequence $F_0\rhdal\ldots\rhdal F_n$,
in a way compatible with Definition \ref{assocredseqdefi}, relying on associated reduction sequences.
Recall the observation that given any reduction sequence $F_0\rhdal\ldots\rhdal F_n$ of terms in $\Tpr$, we may prepend 
a sequence of terms starting with a term in $\T$ which reduces to $F_0$ in at most 
$\lh (F_0)$-many steps and merely involves $(R)$-redices as working redices. We may therefore assume that $F_0\in\T$.
   
\begin{lem}\label{alphaweightlem}
The relation $\wt$ is invariant modulo $\al$-congruence. 
\end{lem}
\proof Straightforward.\qed

The above lemma justifies to consider nice reduction sequences without loss of generality.
The next definition relates derivations to substitution.

\begin{defi}\label{wghtsubstdefi} Let $A,D\in\Tpr$ be such that $D$ is substitutable for $Y$ in $A$, 
i.e.\ $\BV(A)\cap\FV(D)=\emptyset$. For any fixed derivations of $\wt(A,n_A)$ and $\wt(D,n_D)$ 
we define a canonical derivation of $\wt(A\{Y:=D\},n)$ with suitable $n$.  
The definition proceeds by induction along the inductive definition of derivation of $\wt(A,n_A)$.
\begin{enumerate}[(1)]
\item $A= Y$. Then choose $n:=n_D$ and the derivation of $\wt(D,n_D)$.
\item $Y\not\in\FV(A)$. Then choose $n:=n_A$ and the derivation of $\wt(A,n_A)$.
\item $A= BC$ with derivations of $\wt(B,n_B)$ and $\wt(C,n_C)$, and $n_A=n_B+n_C$. 
The canonical derivation of $\wt((BC)\{Y:=D\},n)$ is then assembled from the already defined canonical derivations
$\wt(B\{Y:=D\},n_1)$ and $\wt(C\{Y:=D\},n_2)$, setting $n:=n_1+n_2$.
\item $A=\R^t$ with a derivation of $\wt(t,n_t)$ and $n_A=1+n_t$. 
We have the canonical derivation of $\wt(t\{Y:=D\},n_1)$, from which, setting $n:=1+n_1$,
we define the canonical derivation of $\wt(A\{Y:=D\},n)$.  
\item $A=\lam X.B$, where $X\not= Y$, with a sequence 
$B_0\rhdal\ldots\rhdal B_m= B$ such that $X$ does not occur in any subterm of $B_0$ of a form $\R^t$, 
and derivations of $\wt(B_i,n_{B_i})$ for $i=0,\ldots,m$, and 
$n_A=1+n_{B_0}+\ldots+n_{B_m}$. By the previous lemma we may assume that $D$ is
substitutable for $Y$ in every $B_i$, $i=0,\ldots,m$. 
We already have the canonical derivation of each $\wt(B_i\{Y:=D\},n_i)$. By
assumption, $X$ does not occur in $D$, hence $X$ does not occur in any subterm of
$B_0\{Y:=D\}$ of a form $\R^t$. We have \[B_0\{Y:=D\}\rhdal\ldots\rhdal B_m\{Y:=D\}= B\{Y:=D\}\]
and assembly the canonical derivation of $\wt(A\{Y:=D\},n)$ for $n:=1+n_0+\ldots+n_m$.
\end{enumerate}
\end{defi}

\begin{defi}\label{redseqwdef} We give an inductive definition of canonical weight assignments along reduction sequences.
Suppose $F_0\rhdal\ldots\rhdal F_n$ is a $\lbrpr$-reduction sequence with $F_0\in\T$ and let $p$ be a node in $F_n$ with label $A$.
Assume that derivations of $\wt(B,n_B)$ have been specified for all terms $B$ labeling nodes in the parse trees of $F_i$
for each $i<n$ and to nodes descending from $p$ in the parse tree of $F_n$.
In case $A$ is a variable or constant we are done with the unique derivation of $\wt(A,1)$, 
and if $A= BC$, we have derivations of $\wt(B,n_B)$ and $\wt(C,n_C)$ 
for the labels $B$ and $C$ of the direct child nodes of $p$ and accordingly choose the canonical derivation of $\wt(A,n_B+n_C)$.
The interesting case is where $A=\lam X.G$. Here, let us assume that $F_0\rhdal\ldots\rhdal F_n$ is nice w.r.t.\ $p$, justified 
by Lemma \ref{alphaweightlem}. 
Let $G_0\rhdal\ldots\rhdal G_m$, where $G_m= G$, be the associated reduction sequence for $p$ in $F_0\rhdal\ldots\rhdal F_n$ 
according to Definition \ref{assocredseqdefi}.   
Definition \ref{wghtsubstdefi} then yields canonical derivations $\wt(G_i,n_i)$ for each $i$, from which we assembly the
canonical derivation of $\wt(A,1+n_0+\ldots+n_m)$.
\Fin
\end{defi}

We now see that we may give a non-unique assignment to terms of $\Tpr$ inductively along a definition similar to Definition \ref{wdefi}.
As illustrated in the case of derivations, our assignment method will be unique with respect to assignment derivations.
We can then conveniently prove properties of the assignment method by induction along assignment derivations, cf.\ Definition \ref{wghtsubstdefi}. As illustrated in Definition \ref{redseqwdef} we will have canonical 
assignment derivations in the context of reduction sequences. The assignment method will thus easily be seen to be constructive.

The central result obtained in this paper will be a constructive procedure that, given terms $A,B\in\Tpr$ such that $A\rhd B$ and given an assignment in form of a natural number $a$ to $A$ together with its {\it derivation}, outputs the derivation of an assignment $b\in\Nat$ to $B$ such that $a>b$.          
The uniformity of the procedure then allows for a derivation lengths classification of $\lbrpr$ and hence also of $\lbr$.

\section{Ordinal Terms and Vectors}\label{ordinalsec}

This section provides the theoretical framework of our assignment
of \emph{ordinal vectors} to terms of $\Tpr$. The computational complexity of the calculus $\lbr$ poses a 
challenge regarding the determination of appropriate upper bounds on the lengths of reduction sequences. 
Ordinal terms containing exponential towers occur, see Definition \ref{boxdefi}, and require precise bookkeeping. 
Due to the presence of recursion
with arguments not explicitly known at the beginning of a reduction sequence, the ordinal $\om$ occurs in our assignments,
leading to ordinal terms below the ordinal $\epsn$ which is the least fixed point of exponentiation to base $\om$.
Bookkeeping during the computation of upper bounds will require a sufficiently expressive term algebra, and
using ordinal vectors will help keeping track of exponential (sub-)terms of a certain height. Vectors allow us
to build up ordinal terms from the upper components down, see Definitions \ref{bclassdefi}, \ref{cclassdefi},
\ref{boxdefi}, \ref{dopxdefi}. The $\dop$-operator will require decomposition of ordinal terms, which is facilitated
by the concept of ordinal vectors. 
The starting component is determined by the type level of the term
for which the assignment is being defined, cf.\ Subsection \ref{typelevelsubsec}. 
This approach was designed in \cite{H70} and is used here with some modifications. The innovation, first used by Weiermann
in a treatment of the combinatory logic version of $\GT$, cf.\ \cite{W98}, concerns the $0$-th component.
The $0$th component of the vectors generated will in general result from a collapsing operation below $\om$ and save
an $\om$-power when compared to \cite{H70}\footnote{In \cite{H70}, the $0$-th component introduces another $\omega$-power, 
whereas our modification (essentially) collapses the component $1$.}, see Definition \ref{boxdefi}, 
cf.\ also the difference between \cite{W97} and \cite{W98}.
In order to treat arbitrary $\R$-reductions, where the recursion argument has not yet been reduced to a numeral and
might even contain variables, we feed collapsed terms back into the process of vector generation, see Definition \ref{assigndefi}. 
The $0$th component of a vector assigned to such a recursion argument, say $t$, turns out to be both an upper bound
for the height of the reduction tree of $t$ as well as the {\it value} of $t$ itself, see the proof of Corollary \ref{cortwo}.
Clearly, it is crucial that during a reduction of $t$ (by means of the $\xi$-rule) we obtain a strictly descending chain of ordinal 
terms below $\omega$ in the $0$-th component of the respectively assigned ordinal vectors, while the values of the terms in the 
reduction sequence remain constant. The information on the derivation length of $t$ is therefore needed for the
assignment to the term $\R^t$, and comes genuinely as a natural number which in turn is provided by the collapse in the $0$-th
component of the vector assigned to $t$.
In summary, collapsing plays the essential role in the treatment of unrestricted recursion,
and a more modular procedure, where collapsing is applied in a separate step after the assignment of ordinal terms 
would be at the expense of sharp upper bounds (cf.\ \cite{H70} and \cite{W97}).

We develop an autonomous
theory of ordinal terms and vectors and give an interpretation of closed
ordinal terms as ordinals below $\epsn$. Due to the presence of variables in $\Tpr$ 
we introduce ordinal variables and a notion of comparison on the ordinal terms containing variables.
We adopt most of the terminology and conventions introduced
in Section 2 of \cite{H70}, see in particular the Introduction to Section
2 there. However, knowledge of \cite{H70} is not a prerequisite. The main new ingredient in this paper is the concept of {\it norm}
of an ordinal term and Weiermann's collapsing function $\psi$ which is defined using norms of ordinal notations, see below.

\subsection{Ordinal Terms}\label{ordtermsubsec}

The basic expressions of our assignment are ordinal terms as introduced syntactically in the following definition.
The interpretation of ordinal terms is explained in the remainder of this subsection, together with some facts from ordinal theory.
First of all, let us adopt the following convention regarding ordinal variables. 
 
\noindent {\bf Convention.} Let $\otvar$ be a set of fresh variables, called {\it ordinal variables}.
We assume the existence of a function mapping each typed variable $\xsi$ from $\Var$ to a sequence $(x_0,\ldots,x_{\lv(\si)})$
of pairwise distinct ordinal variables, in such a way that each ordinal variable belongs to exactly one such sequence.
We will sometimes explicitly indicate the type by writing, e.g., $\xsii$ for $x_i$ in the above setting.

\begin{defi}\label{ottermdefi}
The set $\ot$ of ordinal terms is defined inductively as follows:
\begin{iteMize}{$\bullet$}
\item $\otvar\subseteq\ot$.
\item $0,1,\om \in \ot$.
\item If $f,g \in \ot$, then $f+g,\,2^f \cdot g,\,\psiterm{f}{g} \in \ot$.
\end{iteMize}
We call $h\in\ot$ \emph{closed} if it does not contain any variable and  {\it $x$-free} if none of the variables $x_i$ occurs in $h$. The notion of parse tree for ordinal terms is 
clear from the above inductive definition, that is, the immediate subterms of $f+g$, $2^f\cdot g$, and $\psiterm{f}{g}$ are $f$ and $g$. If $g$ in $\psiterm{f}{g}$ is itself
a sum, we will sometimes drop parentheses. Also, if $h\in\ot$ and $n\in\Nat$ we sometimes write $nh$ or $n\cdot h$ in order to denote the $n$-fold summation of $h$.\Fin
\end{defi}

As mentioned above we will interpret closed $\ot$-terms as ordinal numbers below the ordinal $\epsn$, the least fixed point of exponentiation to base $\om$ and hence the proof-theoretic ordinal of Peano arithmetic, as was shown by Gentzen.  
The interpretation of $\ot$-terms will make use of the 
natural sum and product of ordinals, exponentiation to base $2$, as well as the $\psi$-function which was introduced in \cite{W96}. 
For the readers' convenience we are going to recall these ordinal functions, starting with the natural sum $\oplus$ and the natural product $\otimes$ of ordinals, also 
called Hessenberg sum and product, respectively, as well as the exponentiation to bases $2$ and $\om$, $\om$ denoting the least infinite ordinal. 
The natural sum of $\al$ and $0$ agrees with ordinal addition, $\al\oplus 0=0\oplus\al=\al$, 
for the natural product of $\al$ and $0$ we have $\al\otimes 0=0\otimes\al=0$ in agreement with ordinal multiplication. Now let 
\[\al=\om^{\ga_0}+\ldots+\om^{\ga_m}>\ga_0\ge\ldots\ge\ga_m,\; m\ge 0,\] 
and 
\[\be=\om^{\ga_{m+1}}+\ldots+\om^{\ga_n}>\ga_{m+1}\ge\ldots\ge\ga_n,\; n\ge m+1,\] 
be the Cantor normal form representations of non-zero ordinals $\al,\be$ below $\epsn$, 
where $+$ is ordinal addition and $\xi\mapsto\om^\xi$ enumerates the
non-zero ordinals which are closed under ordinal addition, also called additive principal numbers
\footnote{Cf.\ \cite{P09} for a comprehensive introduction to the basics of ordinal arithmetic, 
the proof theory of Peano arithmetic and further advanced topics of proof theory.}.
The natural sum of $\al$ and $\be$ is then defined by
\[\al\oplus\be=\om^{\ga_{\pi(0)}}+\ldots+\om^{\ga_{\pi(n)}}\]  
where $\pi$ is a permutation of $\{0,\ldots,n\}$ such that $\ga_{\pi(0)}\ge\ldots\ge\ga_{\pi(n)}$. 
The natural product of $\al$ and $\be$ is then defined by
\[\al\otimes\be=(\om^{\ga_0\oplus\ga_{m+1}}\oplus\ldots\oplus\om^{\ga_0\oplus\ga_n})\oplus\ldots\oplus(\om^{\ga_m\oplus\ga_{m+1}}\oplus\ldots\oplus\om^{\ga_m\oplus\ga_n}).\]
Exponentiation to base $2$ is characterized by 
\[2^\al=\om^{\al_0}\cdot2^n\] where $\al = \om\cdot\al_0 + n$
and $n < \om$ (cf. for example \cite{S77}). In other words: If $\al$ is the $n$-th successor of the $\al_0$-th limit ordinal,
then $2^\al$ is $2^n$ times the $\al_0$-th additive principal number.

We define the function $\psi : \epsn \to \om$, which will be used to interpret
$\psi$-terms of $\ot$, exactly as in \cite{W96, W97, W98}, where the interested reader can find a detailed 
explanation of the collapsing mechanism and its relation to the theory of subrecursive functions and ordinal recursion.
An abstract exposition of the underlying concepts can be found in \cite{BCW94}, where among other results a comparison of 
$\psi$ with the classical Hardy functions is given. The Hardy functions provide a fine scale that allows for the
comparison of provably recursive functions of fragments of Peano Arithmetic or even the distinction in levels of the 
Grzegorczyk hierarchy. 
By directly assigning $\psi$-terms to terms of $\T$ we can compare the ``run-time'' of different terms in $\T$,
varying in the occurrence of, say, $\R$-functionals of various types. The assignment using $\psi$ resolves nested occurrences
of recursion in terms of $\T$ into un-nested recursion along corresponding ordinal lengths.

Let $\Phi : \om \to \om$ be a sufficiently fast growing number theoretic function, for
example the function $x\mapsto F_5(x + 100)$ where $F_0(x) := 2^x$ and
$F_{n+1}(x):=F_n^{x+1}(x)$. There is leeway in the choice of $\Phi$, however, it is essential that $\Phi$ is bounded
by some $F_k$, $k<\om$, which guarantees the primitive recursiveness of $\Phi$.
Let further the norm function $\no : \epsn \to \om$ be defined by
$\no(0) := 0$ and \[\no(\al):=n + \no(\al_1) + \ldots + \no(\al_n)\]
for $\al = \om^{\al_1} + \ldots + \om^{\al_n} > \al_1 \ge \ldots \ge \al_n$.
This definition of a norm has the convenient property that $\no(k)=k$ for any $k<\om$, 
and we have that for every $m<\om$ the set
\[\{\be\mid\no(\be)\le m\}\]
is \emph{finite}. This property of the norm makes the following definition of the
collapsing function $\psi$ by recursion on $\epsn$ possible:
\[\psi(\al) := \max(\{0\} \cup \{\psi(\be) + 1 \mid \be < \al \,\&\,\no(\be) \le \Phi(\no(\al))\}).\]
This definition can be carried out in $\mathrm{PRA} + \mathrm{PRWO}(\epsn)$, cf.\ \cite{W96,W98}.
We now state two basic propositions concerning the norm and the $\psi$-function (see \cite{W98}).

\begin{prop}\label{propone}
 Let $\al$ and $\be$ be ordinals less than $\epsn$. Then we have
 \begin{enumerate}[\em(1)]
  \item\label{proponei} $\no(\al \oplus \be) = \no(\al) + \no(\be)$.
  \item\label{proponeii} $\no(\al) + \no(\be) - 1 \le \no(\al\otimes\be)
         \le \no(\al) \cdot \no(\be) \mbox{ if } \al \not= 0 \not= \be$.
  \item\label{proponeiii} $\no(\al) \le 2\cdot \no(2^\al)$ and $\no(2^\al) \le 2^{\no(\al)}$. 
 \end{enumerate}
\end{prop}
\proof The proof is given in full detail in the appendix.\qed

\begin{prop}\label{proptwo}
 Let $k<\om$ and ordinals $\al,\,\be<\epsn$ be given. Then we have
 \begin{enumerate}[\em(1)]
  \item\label{proptwoone}
    $k = \psi(k)$, $k \le \psi(\al + k)$, $\no(\al) \le \psi(\al)$, and $\psi(\be) + k \le \psi(\be + k)$.
  \item\label{proptwotwo} $\psi(\al)+\psi(\be) \le \psi(\al\oplus\psi(\be)) \le \psi(\al\oplus\be)$.
  \item\label{proptwothree} $\al < \be \,\&\, \no(\al) \le \Phi(\no(\be)) \imp
         \psi(\al) < \psi(\be)$.
  \item\label{proptwofour} $\al \ge \om \imp \Phi(\no(\al)) < \psi(\al)$. 
 \end{enumerate}
\end{prop}
\proof The proof is given in full detail in the appendix.\qed

Our preparations now enable us to introduce a canonical interpretation
of closed $\ot$-terms as ordinals below $\epsn$. 
This clarifies the notion of the norm $\no(h)$ for closed terms $h\in\ot$
and how closed $\ot$-terms can be compared. 
\begin{defi}\label{otterminterpretedefi}
Closed terms $h\in\ot$ are interpreted canonically, however, $+$ is interpreted by $\oplus$ and $\cdot$ is 
interpreted by $\otimes$.\Fin
\end{defi}
In order to clarify the above definition consider the example of the closed $\ot$-term
\[\psi\left(\om\cdot\left(2^{\om+1}\cdot(\om+0)\right)+2^{2^{\om+1}\cdot\om}\cdot1\right)\]
which is interpreted by the ordinal $\psi(\om\otimes(2^{\om\oplus1}\otimes(\om\oplus0))\oplus2^{2^{\om\oplus1}\otimes\om}\otimes1)=\psi(\om^{\om\cdot2}+\om^3\cdot2)$.

\noindent{\bf Convention.} When working with (closed) $\ot$-terms we will always assume their interpretation by Definition \ref{otterminterpretedefi} and
compare them accordingly.
Therefore, instead of using the symbols $\oplus$ and $\otimes$ we are going to simply use the ordinary symbols $+$ and $\cdot$ in order to refer to 
the natural sum and product, respectively.

\subsection{Ordinal Vectors} 
As in \cite{H70}, a {\it vector of level $n$} is an $n+1$-tuple 
$\vech=\langle h_0,\ldots,h_n\rangle$ where the $h_i$ are ordinal terms,
hence $\vech\in\otlom$.
As there is no danger of ambiguity we write $\lv(\vech)=n$.
For simplicity, we define $h_i$ to be $0$ if $i>\lv(\vech)$. Thus, the sum
$\vech=\vecf+\vecg$ of 
vectors $\vecf,\vecg$ is a vector of level $\max\{\lv(\vecf),\lv(\vecg)\}$  where $h_i=f_i+g_i$. 

The comparison of arbitrary $\ot$-terms and their norms poses the problem of how to interpret or substitute variables. 
We can consider $\ot$-terms and their norms as functions in the variables occurring in them and then define the comparison relation via pointwise domination of functions. 
These functions are viewed as intensional objects and are given by the buildup of the corresponding $\ot$-term. 

Let $X$ be a variable of type $\si$. The {\it variable vector} associated with $X$ is the vector 
\[\vecx:=\langle x_0,\ldots,x_{\lv(\si)}\rangle\]
where $(x_0,\ldots,x_{\lv(\si)})$ is the sequence of ordinal variables associated with $X$ according to the convention at the
beginning of the previous subsection.
We will sometimes explicitly indicate the type by writing, e.g., $\vecxsi$ for $\vecx$ or $\xsii$ for a component $x_i$ of the 
vector $\vecx$ corresponding to $\xsi$. 
By convention we write $\vecx$ for the variable vector associated with $X$, similarly we write $\vecy$ for the variable vector
associated with $Y$, etc.

We will later introduce operations $\bop$ and $\dopx$ (where $\vecx$ ranges over variable vectors) on
vectors that will be applied in our assignment in order to handle application and abstraction, respectively. 
Suitable domains for $\bop$ and $\dopx$, namely classes $\C$ and $\Cx$, 
respectively, will be defined towards the end of this subsection.

{\it Substitution} of variable vectors is defined as follows. For a variable vector $\vecx$ of level $n$ and a vector $\veca$
of ordinal terms of the same level we define the substitution $\{\vecx:=\veca\}$ as the replacement of $x_i$ by $a_i$ for
each $i\le n$. We write $\{\vecx:=\vec{1}\}$ for the replacement of $x_i$ by $1$ for $i\le n$, etc.

The question arises over which domain the substitutions of variable vectors should vary. 
The variable vectors involved are those, of which at least one component occurs in one of the $\ot$-terms 
that are compared or whose norms are compared. We are going to introduce the notion of bounded norm and 
the class $\B$ of vectors characterizing the restrictions of $\Cx$ and $\C$
to closed vectors, i.e.\ vectors whose components are closed ordinal terms.
The restriction of $\B$ to vectors of bounded norm will then serve as the domain of substitutions considered for the 
comparison relation. The operators $\dopx$ will be defined via vectors from $\Cx$ which in general are not of bounded norm, 
which is the reason for not integrating this condition into the definitions of $\B,\Cx,\C$. 

\begin{defi} A closed vector $\vecf\in\otlom$ is of {\it bounded norm} if 
\[\no(f_i)\le \no(f_0)\]
for every $i$.
\end{defi}

\begin{defi}\label{bclassdefi}
We define sets $\Bi \ssq \ot$ for $i<\om$ by simultaneous induction:
\begin{iteMize}{$\bullet$}
\item $1 \in \Bi$ for all $i$.
\item $\om\in\Bi$ for $i\ge 1$.
\item If $f, g \in \Bi$, then $f + g \in \Bi$ for all $i$.
\item If $f \in \Bip$ and $g \in \Bi$, then $2^f \cdot g \in \Bi$ for $i\ge 1$.
\item If $f \in \Be$ and $g \in \Bn$ with\footnote{This condition will be useful later; all $\psi$-terms involved in our assignment will satisfy this condition.} $\no(f)\le F_2(g)$, then 
$\psi(\om \cdot f + g) \in \Bn$.
\item If $h\in\Bn$, then $h\in\Bi$ for $i\ge 1$.
\end{iteMize}


\noindent The class $\B \ssq \otlom$ consists of all vectors $\vech$ such that $h_i\in\Bi$ for all $i\le\lv(\vech)$.
\Fin
\end{defi}
Notice that $0$ does not occur in the parse tree of any term in any $\Bi$.
Terms $h\in\Bn$ cannot be of a shape $2^f\cdot g$, and they
satisfy $h<\om$, which plays a crucial role in this paper as the $0$-th component of a vector assigned to a term in $\Tpr$ is intended to yield an upper bound on the height of the term's reduction tree. 

\begin{defi}\label{comparisondefi} Let $f,g\in\ot$. The relation $f\prec g$ holds if and only if $\chi(f)<\chi(g)$ for all substitutions 
$\chi$ satisfying the following conditions:
\begin{enumerate}[(1)]
\item The domain of $\chi$ is the set of elements of all variable vectors $\vecx$ such that at least one element of $\vecx$
occurs in $f$ or in $g$,
\item For each such $\vecx$ the vector $\chi(\vecx):=\langle \chi(x_0),\ldots,\chi(x_n)\rangle$, where $n={\lv(\vecx)}$, 
is an element of $\B$ of bounded norm.
\end{enumerate} 
The relation $\preceq$ on $\ot$-terms is defined similarly, as well as (extensional) equality $=$.  

The comparison of $\no(f)$ with $\no(g)$ or some term $h\in\Bn$ is defined accordingly, using the same symbols $\prec$, $\preceq$, 
and $=$.

We are going to use the relations $\prec$, $\preceq$, and $=$ also when comparing expressions containing the functions $\no$ and $F_i$, $i<\om$.
Clearly, for expressions of the form $F_i(f)$ to make sense we must have $f\prec\om$. 
\Fin
\end{defi}
Notice that this definition implies that 
\[\no(h)=h\mbox{ for every }h\in\ot\mbox{ such that }h\prec\om,\]
\[0\prec\no(x_i)\preceq x_0\prec\om,\quad\mbox{ and }\quad h\prec\epsn\mbox{ for every }h\in\ot.\]
Propositions \ref{propone} and \ref{proptwo} now generalize to $\ot$-terms and their norms using the above generalized comparison relations. 

\begin{defi}\label{precvecdefi}
Let $\vecf,\vecg\in\otlom$. We define 
\[\vecf \prec \vecg \;\::\aeq\;\: f_0 \prec g_0
  \;\:\&\;\: \fa{i>0} f_i \preceq g_i\]
and 
\[\vecf \preceq \vecg \;\::\aeq\;\: \fa{i} f_i \preceq g_i.\] 
Equality is componentwise equality in the sense of the previous definition.

\medskip
\noindent $\vecf$ is of bounded norm, if
\[\no(f_i)\preceq \no(f_0)\]
for every $i$.
\Fin
\end{defi}
Note that all comparison relations defined in this subsection are transitive, but in general not total.
We conjecture that there is a way to make the comparison relations introduced here effective, however, as we do not need 
such effectiveness in order to achieve our results, we stay with the above elegant comparison notion.

We now adapt the class $\C$, introduced in \cite{H70}
as domain of the operators $\dop^r$. Our version is variable-specific, depending on the abstraction variable, 
so we will introduce classes $\Cx$ serving as domains for $\dopx$.
The classes $\Ci$ and $\C$ defined here comprise the union of all $\Cxi$ and 
$\Cx$, respectively, and will become the general domain of ordinal
vectors used in this article. We will make use of $\C$-vectors which are not of bounded norm, namely when 
defining $\dopx$ in terms of partial operators $\dopxsii$. However, the vectors we are going to assign to terms
of $\Tpr$ will always be $\C$-vectors of bounded norm.  

\begin{defi}\label{cclassdefi}
For every $\vecxsi\in\otlom$, corresponding to some variable $\xsi\in\Var$, we define sets $\Cxi \ssq \ot$ for $i<\om$ by simultaneous induction:
\begin{iteMize}{$\bullet$}
\item $1 \in \Cxi$ for all $i$.
\item $\om\in\Cxi$ for $i\ge 1$.
\item $\yrhoi \in \Cxi$ for $i\le \lv(\rho)$ where $\yrho\in\Var$.
\item If $f, g \in \Cxi$, then $f + g \in \Cxi$ for all $i$.
\item If $f \in \Cxip$ and $g \in \Cxi$, then $2^f \cdot g \in \Cxi$ for $i\ge 1$.
\item If $f \in \Cxe$ and $g \in \Cxn$ with $\no(f)\preceq F_2(g)$, then 
$\psi(\om \cdot f + g) \in \Cxn$.
\item If $h\in\Cxn$ is $x$-free, then $h\in\Cxi$ for $i\ge 1$.
\end{iteMize}

\medskip

\noindent The class $\Cx \ssq \otlom$ is defined to consist of all $\vech$ such that $h_i\in\Cxi$ for all $i\le\lv(\vech)$.
\medskip

\noindent Classes $\Ci$ and $\C$ are defined in the same way as $\Cxi$ and $\Cx$ with the only difference that the condition
of being $x$-free in the last clause defining the classes $\Cxi$ is dropped. 
\Fin
\end{defi}
It is easy to see that the sets of closed $\Cx$-vectors, closed $\C$-vectors, and $\B$-vectors coincide. 
Notice that if $h\in\Cxi$, then it does not contain any variable $x_j$ such that $j<i$, cf.\ Lemma 2.7 of \cite{H70}.
Notice further that the class $\Cx$ is closed under substitution with 
$x$-free $\Cx$-vectors, cf.\ Lemma 2.9 of \cite{H70}. We obviously have $\Cxi\ssq\Ci$ and $\Cx\ssq\C$, and $\C$ is closed under substitution
with $\C$-vectors.

\subsection{\boldmath The Operator $\bop$\unboldmath}\label{boxsubsec}

We now define the operator $\bop$ and show its basic
properties. The definition originates from Howard's \cite{H70} and was used
by Sch\"utte, see \cite{S77}, for an analysis of $\GTCL$ which in turn later served as starting point for 
Weiermann's \cite{W98} with a refinement for vector level $0$ that enabled a derivation lengths classification of G\"odel's T in the combinatory
logic variant. The modification of
the $0$-th level has two important effects: firstly, it saves one $\om$-power,
and secondly, by using the collapsing function $\psi$ we obtain an assignment
of natural numbers to terms in $\Tpr$ instead of ordinal terms below $\epsn$.

\begin{defi}\label{boxdefi}
Let $\vecf, \vecg\in\otlom$ be such that $m:=\lv(\vecf)>\lv(\vecg)=:n$.
We define\footnote{Compared to \cite{H70} we have chosen an asymmetric (non-commutative) definition in order to 
more directly fit the intended application of $\bop$ and facilitate the syntactic fit with our version of $\C$, e.g.\
avoiding occurrences of $0$ in the parse trees of components of $\vecf\bopd\vecg$ for $\vecf,\vecg\in\C$.}
\[ (\vecf \bopd \vecg)_i := \left\{ \begin{array}{l@{\:\mbox{ if }\:}l}
   \psi(\om \cdot (\vecf \bopd \vecg)_1
        + f_0 + g_0 + n) & i = 0 \\[2mm]
   2^{(\vecf \bopd\vecg)_{i+1}} \cdot (f_i + g_i) &
   1 \le i \le n \\[2mm]
   f_i & n < i \le m
   \end{array} \right. \]
to obtain the vector $\vecf\bopd\vecg$ of level $m$.
For $i\le m$ we define $\vecf\bopd\vecg\!\restriction_i$ to be the 
vector of level $i$ whose components are $(\vecf\bopd\vecg)_j$
for $j=0,\ldots,i$.
\Fin
\end{defi}

\begin{lem}\label{boxnormlem} Let $\vecf,\vecg\in\otlom$ be such that $m:=\lv(\vecf)>\lv(\vecg)=:n$ and let $k\in[1,m]$.
Suppose expressions $f, g\prec\om$ satisfy $\no(f_i)\preceq f$ for $k\le i\le m$ and $\no(g_i)\preceq g$ for $k\le i\le n$
(in the sense explained at the end of Definition \ref{comparisondefi}). Then we have
\[\no\left((\vecf\bopd\vecg)_i\right)\preceq F_2(f+g+n)\]
for $k\le i\le m$.
\end{lem}
\proof The proof is given in the appendix.\qed

By the above lemma it follows that $\B$ is closed under $\bop$,
as well as are $\Cx$ and $\C$, cf.\ Lemma 2.8 of \cite{H70}.
If $f_1+g_1\succ 0$ in the case $\lv(\vecg)>0$,
then due to Proposition \ref{proptwo}, Part \ref{proptwofour}, 
$\vecf\bopd\vecg$ is of bounded norm. 
We therefore have the following 

\begin{cor}\label{boxnormcor}
The restriction of $\bop$ to $\C$-vectors maintains bounded norm.
\qed
\end{cor}

We now provide several lemmas which establish crucial properties of $\bop$ and 
are adapted from \cite{H70}, extended to work for component $0$.
\begin{lem}\label{boxone}
Let $\veca,\vecb\in\C$ such that $m:=\lv(\veca)>\lv(\vecb)$.
Then we have 
\[\veca+\vecb\preceq\veca\bopd\vecb.\]
\end{lem}
\proof By induction on $m\minusp i$ it is easily shown that $a_i+b_i\preceq(\veca\bopd\vecb)_i$.
\qed

\begin{lem}\label{boxtwo} Let $\veca,\vecb,\vecc\in\C$.
\begin{enumerate}[\em(1)]
\item\label{boxtwoi} Suppose $m:=\lv(\veca)=\lv(\vecb)>\lv(\vecc)=:n$ and let $i\in[1,m]$.

If $a_j\preceq b_j$ for $i\le j\le m$, then \[(\veca\bopd\vecc)_i\preceq(\vecb\bopd\vecc)_i,\]
where ``$\prec$'' holds, if additionally $a_k\prec b_k$ for some $k\in[i,n+1]$.

If the vectors $\veca$, $\vecb$, and $\vecc$ are of bounded norm and $\veca\preceq\vecb$, then we have
\[\veca\bopd\vecc\preceq\vecb\bopd\vecc,\]
where ``$\prec$'' holds, if $\veca\prec\vecb$.
\item\label{boxtwoii} Suppose $\lv(\veca)>\lv(\vecb),\lv(\vecc)=:n$ and let $i\in[1,n]$.

If $b_j\preceq c_j$ for $i\le j\le n$, then \[(\veca\bopd\vecb)_i\preceq(\veca\bopd\vecc)_i,\]
where ``$\prec$'' holds, if additionally $b_k\prec c_k$ for some $k\in[i,n]$.

If the vectors $\veca$, $\vecb$, and $\vecc$ are of bounded norm and $\vecb\preceq\vecc$, then we have
\[\veca\bopd\vecb\preceq\veca\bopd\vecc,\]
where ``$\prec$'' holds, if $\vecb\prec\vecc$.
\end{enumerate}
\end{lem}
\proof The proof of part \ref{boxtwoi} is by straightforward induction on $m\minusp i$ for the claims concerning components $i$,
$1\le i\le m$. The claim for component $0$ then follows by straightforward application of Proposition \ref{proptwo}, part 
\ref{proptwothree}, and Lemma \ref{boxnormlem}, whose assumptions are satisfied since we have assumed bounded norms.
The proof of part \ref{boxtwoii} is analogous.
\qed

\begin{lem}\label{boxthree}
Let $\veca,\vecb,\vecc,\vecd\in\C$ be such that $m:=\lv(\veca)=\lv(\vecb)=\lv(\vecc)>\lv(\vecd)=:n$.
\begin{enumerate}[\em(1)]
\item\label{boxthreei} If $a_i + b_i\preceq c_i$ for $1\le i \le n+1$, then 
\[(\veca\bopd\vecd)_i+(\vecb\bopd\vecd)_i\prec(\vecc\bopd\vecd)_i\]
for $1\le i\le n$.
\item\label{boxthreeii} If $\no(a_i)\preceq\no(b_i)$ for $1\le i\le m$, then 
\[\no((\veca\bopd\vecd)_i)\preceq F_3\left(\no((\vecb\bopd\vecd)_i)+n\right)\]
for $1\le i\le m$.
\item\label{boxthreeiii} If $a_i+b_i\prec c_i$ and $\no(a_i),\no(b_i)\preceq\no(c_i)$ for $i\le m$, then
\[(\veca\bopd\vecd)_i+(\vecb\bopd\vecd)_i\prec(\vecc\bopd\vecd)_i\]
for $i\le m$.
\end{enumerate}
\end{lem}
\proof For the detailed proof the reader is referred to the appendix.
\qed

The next lemma is an adaptation of the crucial Lemma 2.6 of \cite{H70}.
It is the key to the treatment of the combinatorial complexity of the combinator $\mathsf{S}$, cf.\ \cite{S77,W98},
in part of the recursor $\R$, and, when combined with the operator $\dop$, of $\be$-reduction.
Regarding the latter property which is essential in Howard's approach, notice the correspondence of
the factor $2$ occurring in Definition \ref{dopxdefi} in the case of $\dopxsii h$ where $h\equiv2^f\cdot g$ is not $x$-free 
(this factor is in fact only necessary for the highest vector component),
with the factor $2$ in the assumption $2a_{n+1}+b_{n+1}\prec c_{n+1}$ of the following lemma. We regard this correspondence 
as crucial in order to understand why the operators $\bopd$ and $\dopx$ model $\be$-reduction. Consider as instructive example
$\be$-reduction of terms of the form $(\lam X.AB)D$.

\begin{lem}\label{boxfour}
Let $\veca,\vecb,\vecc,\vecd\in\C$ and $n\in\Nat$ be such that $\lv(\veca)=\lv(\vecb)=\lv(\vecc)=n+1$ and $\lv(\vecd)=n$.
If $a_i+b_i\prec c_i$ for $1\le i\le n$ and $2a_{n+1}+b_{n+1}\prec c_{n+1}$, then setting
\[\vece:=(\veca\bopd\vecd)\bopd(\vecb\bopd\vecd\!\restriction_n)\]
we have
\[2e_i\prec(\vecc\bopd\vecd)_i\]
for $1\le i\le n+1$.
\end{lem}
\proof 
See the appendix for a proof in full detail.
\qed

For the treatment of $\R$-reductions we will need estimations of norms of the type stated in the following lemma.
\begin{lem}\label{boxextnormlem}
Let $\veca,\vecb,\vecc,\vecd\in\C$ be of bounded norm and $n\in\Nat$ such that $\lv(\veca)=\lv(\vecc)=n+1>\lv(\vecb),\lv(\vecd)$.
Setting  
\[\vece:=(\veca\bopd\vecb)\bopd(\vecc\bopd\vecd\!\restriction_n)\]
we have
\[\no(e_i)\prec F_3(a_0+b_0+c_0+d_0+n)\]
for $1\le i\le n+1$.
\end{lem}
\proof The proof is given in the appendix.
\qed

\subsection{\boldmath The Operators $\dopx$\unboldmath}\label{dopsec}
Here we introduce our refinement of the operators $\dop^r$, see \cite{H70}, 
which provide the key to appropriate assignments of ordinal vectors to abstraction terms in order to allow for
the treatment of $\be$-contraction. 
Our modification of  $\dop$ essentially concerns vector level zero, which ranges over terms for natural numbers instead of
ordinals below $\epsn$ as in the original version. This is made possible by application of the collapsing function $\psi$. 
Our refinement is formulated on the basis
of the $\Cx$-classes introduced earlier in order to make the treatment
of general $\R$-reductions possible. 

\begin{defi}\label{dopxdefi}
In order to define $\dopx\::\:\Cx\to\Cx$ let $\vech\in\Cx$ be of level $m:=\lv(\vech)$ and set $n:=\lv(\vecx)+1$.
$\dopx\vech$ is then a vector of level $l:=\max\{n,m\}$, defined componentwise by
\[(\dopx\vech)_j := \left\{
  \begin{array}{l@{\:\mbox{ if }\:}l}
     \lx(\vech)\cdot(\dopxsin h_0)_0 & j = 0 \\[2mm]
     \sum^m_{i=0}(\dopxsii h_i)_j & 0 < j \le n \\[2mm]
     h_j & n < j \le l
  \end{array} \right. \]
where $\lx$ will be defined below and the $\dopxsii:\Cxi\to\Cx$ are defined recursively as follows. 
Let $h\in\Cxi$. $\dopxsii h$ is a $\Cx$-vector of level $n$, defined componentwise as follows.
If $h$ is $x$-free we set
\begin{iteMize}{$\bullet$}\item[] \hfill\\
         $(\dopxsii h)_j :=
           \left\{\begin{array}{l@{\:\mbox{ if }\:}l}
           1 & i\not=j\le n \\
           h+1 & i=j\le n.
           \end{array}\right.$
\end{iteMize}
The following cases apply if $h$ is not $x$-free.
\begin{iteMize}{$\bullet$}
   \item $h \equiv \xsii$:\\[2mm]
         $\dopxsii h:= \vec{1}$.\vspace{2mm}
   \item $h\equiv f+g$ where $f, g \in \Cxi$:\\[2mm]
         $\dopxsii h:= \dopxsii f + \dopxsii g + \vec{1}$.\vspace{2mm}
   \item $h\equiv2^f\cdot g$ where $f\in\Cxip, g\in\Cxi$, and $i>0$:\\[2mm]
         $\dopxsii h := 2\dopxsiip f + \dopxsii g +\vec{1}$.\vspace{2mm}
   \item $h \equiv \psiterm{f}{g}$
         where $f\in\Cxe, g\in\Cxn$ and $i=0$:\\[2mm]
         $(\dopxsii h)_j :=
           \left \{ \begin{array}{l@{\:\mbox{ if }\:}l}
           \psiterm{f\{\vecx:= \vec{1}\}}
                   {(\dopxsin g)_0} & j=0 \\[1mm]
           (\dopxsie f)_j+(\dopxsin g)_j & 1\le j\le n.
           \end{array}\right.$
\end{iteMize}
The norm controlling factor $\lx(\vech)\in\Nat$ is defined by
\[\lx(\vech):=2^n\cdot\sum_{i=0}^m\lhx(h_i)\]
where the auxiliary $\lhx(h)$ for $h\in\ot$ is defined by
\begin{iteMize}{$\bullet$}
\item $\lhx(h):=1$ if $h$ is $x$-free or $h\equiv x_i$ for some $i$.
\item $\lhx(h):=\lhx(f)+\lhx(g)+1$ if $h$ is not $x$-free and either of a form $f+g$ or $\psiterm{f}{g}$.
\item $\lhx(h):=2\lhx(f)+\lhx(g)+1$ if $h$ is not $x$-free and of a form $2^f\cdot g$. \Fin
\end{iteMize}
\end{defi}
Notice that $\dopx\vech$ does not contain any component of $\vecx$. 
In order to see that the above definition is sound, we have to verify that the vectors $\dopxsii h$ are indeed $\Cx$-vectors. 
We have the following 
\begin{lem}\label{crucialestimlem} For $h\in\Cxn$ we have
\[h\{\vecx:=\vec{1}\}\prec(\dopxsin h)_0.\]
\end{lem}
\proof The proof is by induction on the buildup of $h$. The interesting case is where $h$ is of the form $\psiterm{f}{g}$ and not $x$-free.
We then use the i.h.\ for $g$, obtaining
\begin{eqnarray*}
h\{\vecx:=\vec{1}\}&\equiv&\psiterm{f\{\vecx:=\vec{1}\}}{g\{\vecx:=\vec{1}\}}\\
&\prec&\psiterm{f\{\vecx:=\vec{1}\}}{(\dopxsin g)_0}\\
&\equiv&(\dopxsin h)_0.
\end{eqnarray*}\qed

The above lemma shows that for terms $h\equiv\psiterm{f}{g}\in\Cxn$ we have \[\no(f\{\vecx:=\vec{1}\})\preceq F_2((\dopxsin g)_0),\]
using that $\no(f\{\vecx:=\vec{1}\})\preceq F_2(g\{\vecx:=\vec{1}\})$. It is then easy to verify that $\dopxsii h\in\Cx$ for $h\in\Cxi$ and hence
$\dopx\vech\in\Cx$ for $\vech\in\Cx$. 

We call a substitution $\{\vecy:=\vecg\}$ an {\it $x$-free substitution} if $\vecy\not\equiv\vecx$ and $\vecg$ is $x$-free. This notion facilitates an elegant statement of the next lemma, which corresponds to Lemma 2.10 and Corollary of \cite{H70}.
\begin{lem}\label{substdopsilem} The operator $\dopx$ is commutes with $x$-free substitution: 
for $\vecf,\vech\in\Cx$, $\vecf$ $x$-free, and $\vecy\not\equiv\vecx$ we have
\[(\dopx\vech)\{\vecy:=\vecf\}=\dopx(\vech\{\vecy:=\vecf\}).\]
\end{lem}
\proof Notice that $\lhx$ and hence $\lx$ are invariant under $x$-free substitution. It is then straightforward to verify the commutativity of the partial operators $\dopxsii$ with $x$-free
substitution and finally conclude the lemma.
\qed

\begin{lem}\label{dopboundednormlem}
For any variable vector $\vecx$ the operator $\dopx$ preserves bounded norm: for every $\vech\in\Cx$ of bounded
norm $\dopx\vech$ is of bounded norm.
\end{lem}
\proof The lemma is part \ref{doplemthree} of Lemma \ref{doplem} which is stated and proved in the appendix. 
\qed

We conclude this section establishing the interplay of the operators 
$\bop$ and $\dopx$, corresponding to Lemma 2.11 of \cite{H70} 
and its Corollary. 

\begin{lem}\label{interplaylem} Let $\vech\in\Cx$. We have
\[\vech\prec\dopx\vech\bopd\vecx.\]
\end{lem}
\proof The proof is given in the appendix. \qed

\section{Assignment of Ordinal Vectors to Terms}\label{assignsec}
\subsection{Assignment Derivations}
We are now prepared to assign ordinal vectors to terms of $\Tpr$. Recall, for illustrative reasons, Definition \ref{wdefi} and 
the notion of derivation along an inductive definition. Definition \ref{redseqwdef} provides canonical derivations of 
$\wt(F,n_F)$ for every $F\in\Tpr$ along a given reduction sequence. 
With the following inductive definition, which is independent of Subsection \ref{assignsubsec}, 
we refine the notion of derivation towards {\it assignment derivation}, carrying the ordinal vectors 
assigned to terms as labels. 

\begin{defi}\label{assigndefi} We define {\it assignment derivations} inductively for terms $A^\si\in\Tpr$, which assign vectors
$\klamp{A}$ of level $\lv(\si)$ to $A$. The notation $\klamp{A}$ is therefore only determined uniquely in the 
context of a fixed assignment derivation.
\medskip

{\bf Assignment to prime terms of $\Tpr$:}
If $A$ is a variable or constant, then it has a unique assignment
$\klamp{A}$ as defined below, and its assignment derivation is a single-node tree which is labeled with $(A,\klamp{A})$.
\begin{iteMize}{$\bullet$}
\item[] $\klamp{\xsi}:=\vecxsi$.\vspace{2mm}
\item[] $\klamp{0}:=\langle1\rangle$.\vspace{2mm} 
\item[] $\klamp{\Suc}:=\langle1,1\rangle$.\vspace{2mm}
\item[] $\klamp{\Dt}:=\langle1,\ldots,1\rangle$ of level $\lv(\tau)+1$.\vspace{2mm}
\item[] $\klamp{\Rt}:=\langle2,1,\ldots,1,\om\rangle$ of level 
$\lv(\tau)+2$.
\end{iteMize}
\medskip

{\bf Terms formed by application:}
For assignment derivations $\treeR$ of $(B^{\si\tau},\klamp{B})$ and 
$\treeS$ of $(C^\si,\klamp{C})$, the tree with direct subtrees $\treeR$ 
and $\treeS$ is a derivation of $(BC,\klamp{BC})$, where $\klamp{BC}$ is defined by
\[\klamp{BC}:=\klamp{B}\bopd\klamp{C}\!\restriction_{\lv(\tau)}.\] 

\medskip For an assignment derivation $\treeR$ of $(t,\klamp{t})$, the 
tree with direct subtree $\treeR$ is an assignment derivation of 
$(\R^t,\klamp{\R^t})$ where $\klamp{\Rt^t}$ of level $\lv(\tau)+2$ is defined by 
\[\klamp{\Rt^t}:=\langle\klampn{t},1,\ldots,1,\klampn{t}\rangle.\] 
\medskip

{\bf Terms formed by abstraction:} 
For assignment derivations $\treeR_i$ of 
$(G_i,\klamp{G_i})$, $i\le m$, where 
$G_0\rhdal\ldots\rhdal G_m=:G$ such that
$X$ does not occur in any subterm of $G_0$ of a form $\R^t$ and
$\klamp{G_0}\succ\ldots\succ\klamp{G_m}$, the tree with direct subtrees
$\treeR_0,\ldots,\treeR_m$ in this order is an assignment derivation of 
$(\lam X.G,\klamp{\lam X.G})$, the label of its root, where 
\[\klamp{\lam X.G}:=\dopx\klamp{G_0}+\klamp{G_m}\{\vecx:=\vec{1}\}.\]
Whenever a particular assignment derivation is clear from the context of argumentation, we will
use the notation $\klamp{\cdot}$ as if it were an operator returning a unique ordinal vector.

\medskip 
 
For a term $A\in\T$ we define the {\it canonical assignment for $A$} 
by choosing for every subterm of a form $\lam X.G$ the assignment 
$\klamp{\lam X.G}:=\dopx\klamp{G}+\klamp{G}\{\vecx:=\vec{1}\}$.
This results in a unique assignment $\klamp{A}$ to the term $A$.
We call the vector $\veca$ resulting from $\klamp{A}$ by replacing every 
variable by $1$ the {\it closed canonical assignment for $A$}.
\Fin
\end{defi}

It is easy to verify that all vectors assigned to terms are $\C$-vectors of bounded norm, cf.\ 
Corollary \ref{boxnormcor} and Lemma \ref{dopboundednormlem}. Notice also that in case of an assignment
$\klamp{\lam X.G}:=\dopx\klamp{G_0}+\klamp{G_m}\{\vecx:=\vec{1}\}$ the vector $\klamp{G_0}$ even is a $\Cx$-vector, as is
required for the application of the operator $\dopx$. This latter property $\klamp{G_0}\in\Cx$ is guaranteed by the
fact that the variable $X$ corresponding to $\vecx$ does not occur in any subterm of the form $\R^t$ of $G_0$. This is 
crucial for the compatibility of the original treatment of $\be$-reductions with unrestricted $\R$-reductions and is
one of the two reasons why we need this form of non-unique ordinal assignment that depends on the reduction history of
terms in a given reduction sequence, as explained in greater detail in Subsection \ref{assignsubsec}. 
The other reason is the same as in \cite{H70}: the operators $\dopx$ are in general
not monotonically increasing\footnote{Consider for example variables $x,y$ of type $00$,
variables $z,u$ of type $0$ and compute the canonical assignments to $\lam x.((\lam z.x(yz))u)\rhd\lam x.(x(yu))$.
Setting e.g.\ $y:=\lam w^0.w^0$ and $u:=y(yv^0)$ we see that $\dopx$ is not even weakly monotonically increasing.} 
(see also \cite{H70}, p.\ 456) and therefore do not allow for a direct treatment of the unrestricted $\xi$-rule.

\begin{lem}\label{alphaassignlem}
Assignment derivations are invariant modulo $\al$-congruence. 
\end{lem}
\proof Straightforward.\qed

The following lemma corresponds to Lemma 3.1 of \cite{H70}, cf.\ also Definition \ref{wghtsubstdefi}.

\begin{lem}\label{assignsubstlem} The assignment $\klamp{\cdot}$ commutes with substitution. Let $F,H\in\Tpr$ satisfy $\FV(F)\cap\BV(H)=\emptyset$ and
let $Y$ be a variable of the same type as $F$.
Given assignment derivations of $(H,\klamp{H})$ and $(F,\klamp{F})$, there is a canonical assignment derivation
of $(H\{Y:=F\},\klamp{H}\{\vecy:=\klamp{F}\})$, defined straightforwardly in the proof.
\end{lem}
\proof The proof is by induction along the definition of an assignment derivation of $(H,\klamp{H})$.
The interesting case is where $H$ is an abstraction term, say $\lam X.G$, whose assignment 
\[\klamp{H}=\dopx\klamp{G_0}+\klamp{G_m}\{\vecx:=\vec{1}\}\]
is based on assignments $\klamp{G_0},\ldots,\klamp{G_m}$ where $G_0\rhdal\ldots\rhdal G_m= G$ such that
$X$ does not occur in any subterm of $G_0$ of a form $\R^t$
and $\klamp{G_0}\succ\ldots\succ\klamp{G_m}$.

Since the case $Y= X$ is trivial, we assume $Y\not= X$. By assumption we have $X\not\in\FV(F)$, and according to Lemma 
\ref{alphaassignlem} we may further assume w.l.o.g.\ that $\FV(F)\cap\BV(G_0)=\emptyset$. 
The i.h.\ yields assignment derivations of $(G_i\{Y:=F\},\klamp{G_i}\{\vecy:=\klamp{F}\})$ for $i\le m$, and we have
\[\klamp{G_0}\{\vecy:=\klamp{F}\}\succ\ldots\succ\klamp{G_m}\{\vecy:=\klamp{F}\}.\] 
Clearly, 
\[(\klamp{G_m}\{\vecx:=\vec{1}\})\{\vecy:=\klamp{F}\}=(\klamp{G_m}\{\vecy:=\klamp{F}\})\{\vecx:=\vec{1}\},\]
and by Lemma \ref{substdopsilem} we have 
\[(\dopx\klamp{G_0})\{\vecy:=\klamp{F}\}=\dopx(\klamp{G_0}\{\vecy:=\klamp{F}\}).\]
We have \[G_0\{Y:=F\}\rhdal\ldots\rhdal G_m\{Y:=F\},\]
and $X$ does not occur in any subterm of $G_0\{Y:=F\}$ of a form $\R^t$.
The assignment derivation of $(H\{Y:=F\},\klamp{H}\{\vecy:=\klamp{F}\})$ can therefore be assembled from the 
assignment derivations of the $(G_i\{Y:=F\},\klamp{G_i}\{\vecy:=\klamp{F}\})$.
\qed

\begin{defi}\label{assignalgodefi}
By recursion on $\lh(A)$ we define an algorithm which, given terms
$A,B\in\Tpr$ such that $A\rhd B$ and given an assignment derivation for
$A$, returns an assignment derivation for $B$.
\medskip

{\bf Reductions:}
\begin{iteMize}{$\bullet$}
\item \boldmath$(D_0)$, $(D_\Suc)$, $(R)$, 
and $(R^0)$\unboldmath\ are trivial, proceeding in the same way as in the following case.\vspace{2mm}
\item \boldmath$(R^\Suc)$\unboldmath\ $\R^{\Suc t}FG\rhd Ft(\R^tFG)$ with 
$\klamp{\R^{\Suc t}FG}$ given via assignments  
$\klamp{t},\klamp{F},\klamp{G}$.
Then \[\klamp{Ft(\R^tFG)}\] is built up from the same assignments
$\klamp{t},\klamp{F},\klamp{G}$.
\item \boldmath$(\be)$\unboldmath\ $(\lam X.G)H \rhd G\{X:=H\}
\mbox{ where } \BV(\lam X.G) \cap \FV(H) = \emptyset$ with 
$\klamp{(\lam X.G)H}$ given via assignments $\klamp{\lam X.G},\klamp{H}$
where the former is in turn given via assignments 
$\klamp{G_0},\ldots,\klamp{G_m}$ from the assignment derivation of
$\lam X.G$, whence $G_0\rhdal\ldots\rhdal G_m= G$.
By Lemma \ref{assignsubstlem} we obtain an assignment
\[\klamp{G_m}\{\vecx:=\klamp{H}\}\] to the term $G\{X:=H\}$ with the
canonical assignment derivation.       
\end{iteMize}
\medskip

{\bf Rules:}
\begin{iteMize}{$\bullet$}
\item \boldmath$(App_r)$, $(App_l)$, and $(App_R)$\unboldmath\ are 
handled in the same straightforward manner, e.g.\ in the case 
where $\R^s\rhd\R^t$ is derived from $s\rhd t$ and the assignment
$\klamp{\R^s}$ given via an assignment $\klamp{s}$ to $s$, we
let $\klamp{t}$ be the assignment provided by the algorithm
and build $\klamp{\R^t}$ up from $\klamp{t}$.
\item \boldmath$(\xi)$\unboldmath\ $\lam X.F\rhd\lam X.G$ derived from
$F\rhd G$, with $\klamp{\lam X.F}$ given by means of assignments 
$\klamp{F_0},\ldots,\klamp{F_m}$ from the assignment derivation of
$\lam X.F$, whence $F_0\rhdal\ldots\rhdal F_m= F$. We then choose the
assignment 
\[\klamp{\lam X.G}:=\dopx\klamp{F_0} + \klamp{G}\{\vecx:=\vec{1}\},\] 
where $\klamp{G}$ is the assignment to $G$ provided by the algorithm.\Fin
\end{iteMize}\smallskip
\end{defi}

\noindent The soundness of the above definition hinges on the verification 
that in the clause for \boldmath$(\xi)$\unboldmath\ we indeed have
$\klamp{F_m}\succ\klamp{G}$. This is accomplished by our Main Theorem.

\subsection{The Main Theorem.}
\begin{thm}\label{assignthm} For $A,B\in\Tpr$ such that $A\rhd B$ and a given assignment
derivation assigning $\klamp{A}$ to $A$, the assignment $\klamp{B}$ to $B$
that is provided by the algorithm specified in Definition 
\ref{assignalgodefi} satisfies
\[\klamp{A}\succ\klamp{B}.\]  
\end{thm}
\proof The proof is by induction on $\lh(A)$. We use the terminology of Definition \ref{assignalgodefi}.

\medskip
\boldmath$(D_0)$, $(D_\Suc)$, $(R)$, and $(R^0)\,$\unboldmath are handled straightforwardly using Lemma \ref{boxone}.

\medskip 
\boldmath$(R^\Suc)\:$\unboldmath $\R^{\Suc t}FG\rhd Ft(\R^tFG)$ where
$\R\equiv\Rt$ and $n:=\lv(\tau)$. 
Suppose $\klamp{\R^{\Suc t}FG}$ is given via assignments $\klamp{t}$,
$\klamp{F}$, and $\klamp{G}$. We introduce the following abbreviations:
\begin{eqnarray*}
\veca &:=& \klamp{Ft}\\
\vecb &:=& \klamp{\R^tF}\\
\vecc &:=& \klamp{\R^{\Suc t}F}\\
\vecd &:=& \klamp{G}\\
\vece &:=& (\veca \bopd\vecd)\bopd(\vecb\bopd\vecd\!\restriction_n)\\
\vecf &:=& \klamp{F}\\
\vect &:=& \klamp{t}
\end{eqnarray*}
Notice that we have $\veca=\vecf\bopd\vect$ and
\begin{eqnarray*}
\klamp{\R^t}&=&\langle t_0,1,\ldots,1,t_0\rangle\\
&\prec&\langle\psi(\om+t_0+1),1,\ldots,1,\psi(\om+t_0+1)\rangle\\
&=&\klamp{\R^{\Suc t}},
\end{eqnarray*}
hence $b_i=(\klamp{\R^t}\bopd\vecf)_i\prec
(\klamp{\R^{\Suc t}}\bopd\vecf)_i=c_i$ for $i\le n+1$ by Lemma
\ref{boxtwo}, part \ref{boxtwoi}.
We further have
$\klamp{\R^tFG}=\vecb\bopd\vecd\!\restriction_n$,
and 
\[\klamp{Ft(\R^tFG)}=\veca\bopd(\vecb\bopd\vecd\!\restriction_n)\!\restriction_n\prec\vece\]
by part \ref{boxtwoi} of Lemma \ref{boxtwo} since 
$\veca\prec\veca\bopd\vecd$ by Lemma \ref{boxone}.
We obtain 
\[2e_i\prec(\vecc\bopd\vecd)_i\]
for $1\le i\le n+1$ by an application of Lemma \ref{boxfour} whose
assumptions $a_i+b_i\prec c_i$ for $1\le i\le n$ and 
$2a_{n+1}+b_{n+1}\prec c_{n+1}$ are easily verified.
As $\klamp{\R^{\Suc t}FG}=\vecc\bopd\vecd\!\restriction_n$ 
we obtain 
\[2\klamp{Ft(\R^tFG)}_i\prec\klamp{\R^{\Suc t}FG}_i\]
for $1\le i\le n$, and in the case $n>0$ by Lemma \ref{boxone} we thus have
\begin{equation}\label{recequ}
\klamp{Ft(\R^tFG)}_1+f_1+\klamp{\R^tFG}_1\prec\klamp{\R^{\Suc t}FG}_1.
\end{equation}
It remains to prove that 
\begin{equation}\label{zerolevel}
\klamp{Ft(\R^tFG)}_0\prec\klamp{\R^{\Suc t}FG}_0.
\end{equation}
We begin with the following estimation
\begin{eqnarray*}
b_0+f_0+t_0&=&\psiterm{b_1}{t_0+f_0+n+1}+f_0+t_0\\[1mm]
&\preceq&\psiterm{b_1}{2t_0+2f_0+n+1}\\[1mm]
&\prec&\psiterm{c_1}{\psi(\om+t_0+1)+f_0+n+1}\\[1mm]
&=&c_0,
\end{eqnarray*}
which follows by Proposition \ref{proptwo}, part \ref{proptwothree}, since 
$b_1\prec c_1$ and $\no(b_1)\preceq F_2(t_0+f_0+n+1)$ using 
Lemma \ref{boxnormlem}.
In the case $n=0$ it is easy to verify \ref{zerolevel}. 
Let us therefore assume that $n>0$.
Using parts \ref{proptwotwo} and \ref{proptwothree} of Proposition 
\ref{proptwo}, from \ref{recequ} we then obtain 
\begin{eqnarray*}
\klamp{Ft(\R^tFG)}_0&=&
\psiterm{\klamp{Ft(\R^tFG)}_1}{a_0+\klamp{\R^tFG}_0+n}\\[1mm]
&=&\psiterm{\klamp{Ft(\R^tFG)}_1}{\psiterm{f_1}{f_0+t_0}+\\[1mm]
&&\quad\quad\psiterm{\klamp{\R^tFG}_1}{b_0+d_0+n}+n}\\[1mm]
&\preceq&\psiterm{(\klamp{Ft(\R^tFG)}_1+f_1+\klamp{\R^tFG}_1)}{b_0+f_0+d_0+t_0+2n}\\[1mm]
&\prec&\psiterm{\klamp{\R^{\Suc t}FG}_1}{c_0+d_0+n}\\[1mm]
&=&\klamp{\R^{\Suc t}FG}_0,
\end{eqnarray*}
since using Lemma \ref{boxextnormlem} we may estimate
\begin{eqnarray*}
\no(\klamp{Ft(\R^tFG)}_1+f_1+\klamp{\R^tFG}_1)&\preceq&
F_3(f_0+t_0+b_0+d_0+n)+\\[1mm]
&&\quad\quad F_2(f_0+t_0)+F_2(b_0+d_0+n)\\[1mm]
&\prec&\Phi(c_0+d_0+n).
\end{eqnarray*}

\medskip 
\boldmath$(\be)\:$\unboldmath $(\lam X.G)H\rhd G\{X:=H\}$.
With the notations of Definition \ref{assignalgodefi} the vector assigned
to $\lam X.G$ is $\dopx\klamp{G_0}+\klamp{G}\{\vecx:=\vec{1}\}$.
By Lemma \ref{interplaylem} we have 
\[\klamp{G_0}\prec\dopx\klamp{G_0}\bopd\vecx,\]
hence 
\begin{eqnarray*}
\klamp{\lam X.G}\bopd\klamp{H}&\succ&
\dopx\klamp{G_0}\bopd\klamp{H} 
\mbox{ by part \ref{boxtwoi} of Lemma \ref{boxtwo}}\\[1mm]
&\succ&\klamp{G_0}\{\vecx:=\klamp{H}\}.
\end{eqnarray*}
We have $\klamp{G_0}\succ\ldots\succ\klamp{G_m}$,
hence 
\[\klamp{G_0}\{\vecx:=\klamp{H}\}\succ\ldots\succ\klamp{G_m}\{\vecx:=\klamp{H}\},\]
and by Lemma \ref{assignsubstlem}
the $\klamp{G_i}\{\vecx:=\klamp{H}\}$ are vectors assigned to 
$G_i\{X:=H\}$ for $i\le m$. 

\medskip
\boldmath$(App_r)\:$\unboldmath $FH\rhd GH$, derived from $F\rhd G$.
By the i.h.\ we have $\klamp{F}\succ\klamp{G}$, hence part \ref{boxtwoi} 
of Lemma \ref{boxtwo} yields $\klamp{FH}\succ\klamp{GH}$.

\medskip
\boldmath$(App_l)\:$\unboldmath $FG\rhd FH$, derived from $G\rhd H$.
By the i.h.\ we have $\klamp{G}\succ\klamp{H}$, hence part \ref{boxtwoii} of Lemma \ref{boxtwo} yields $\klamp{FG}\succ\klamp{FH}$.

\medskip
\boldmath$(App_R)\:$\unboldmath $\R^s\rhd\R^t$, derived from 
$s\rhd t$. 
By the i.h.\ we have $\klamp{s}\succ\klamp{t}$, so we immediately obtain
$\klamp{\R^s}\succ\klamp{\R^t}$.

\medskip
\boldmath$(\xi)\:$\unboldmath $\lam X.F\rhd\lam X.G$, derived from 
$F\rhd G$.
Then $\klamp{\lam X.F}\succ\klamp{\lam X.G}$ follows directly from the 
i.h.\ which yields $\klamp{F_m}\succ\klamp{G}$.
\qed

\begin{cor}\label{assigncor}
Let $A\in\T$ and $\veca$ be its closed canonical assignment.
Then $a_0\in\Nat$ is an upper bound of the height of the reduction tree
of $A$. We obtain strong normalization for $\lbr$ and $\lbrpr$. 
\end{cor}
\proof Regarding the relationships between reduction sequences of 
$\T$-terms in $\lbr$ and $\Tpr$-terms in $\lbrpr$, 
recall the remarks stated in Subsection \ref{extsubsec}. 
The corollary then follows from Theorem \ref{assignthm}.
\qed

\subsection{Tying in with Subsection \ref{assignsubsec}.}\label{linksubsec}
For illustrative reasons we establish the link of our assignment with Definition \ref{redseqwdef}. 

Let $F_0\rhdal\ldots\rhdal F_n$ be a $\lbrpr$-reduction sequence with $F_0\in\T$ and let $p$ be a node in $F_n$ with label $A$.
Suppose that assignment derivations of $(B,\klamp{B})$ have been specified for all terms $B$ labeled to nodes in the parse trees of 
$F_i$ for $i=0,\ldots,n-1$ and to nodes descending from $p$ in the parse tree of $F_n$.
In case $A$ is a variable or constant we are done with $(A,\klamp{A})$, and if $A^\tau\equiv BC$, we have $(B,\klamp{B})$ and 
$(C,\klamp{C})$ for the labels $B$ and $C$ of the direct child nodes of $p$ and accordingly choose 
$(A,\klamp{B}\bopd\klamp{C}\!\restriction_{\lv(\tau)})$.
The interesting case is where $A=\lam X.G$. We may assume that $F_0\rhdal\ldots\rhdal F_n$ is $p$-nice, according to Lemma
\ref{alphaassignlem}.
Let $(G_0,\ldots,G_m)$, where $G_m= G$, be the associated reduction sequence w.r.t.\ $F_0\rhdal\ldots\rhdal F_n$ and $p$
according to Definition \ref{assocredseqdefi}, so using Lemma \ref{assignsubstlem} we obtain 
assignment derivations labeled with $(G_i,\klamp{G_i})$ for $i\le m$.  
If we can show that 
\[\klamp{G_0}\succ\ldots\succ\klamp{G_m},\]
then we obtain an assignment derivation of 
$(A,\dopx\klamp{G_0}+\klamp{G_m}\{\vecx:=\vec{1}\})$.
We are going to show that the assignments $\klamp{G_1},\ldots,\klamp{G_m}$
are obtained by consecutive application of the algorithm given in 
Definition \ref{assignalgodefi}, starting from $\klamp{G_0}$. Recalling
our description of how the parse tree of $F_{i+1}$ is obtained from
the parse tree of $F_i$ in \ref{basiccons},
we see that for corresponding terms $A$ in the parse tree of $F_i$
and $B$ in the parse tree of $F_{i+1}$ such that $A\rhd B$, the
assignment derivation of $\klamp{B}$ is obtained from the assignment 
derivation of $\klamp{A}$ by the algorithm given in 
Definition \ref{assignalgodefi}. Such terms $A$ and $B$ are either 
the working redex itself and its reduct, corresponding to a $\D$-,
$\R$-, or $\be$-reduction, 
or corresponding terms on the
paths leading from the roots of the parse trees of $F_i$ and $F_{i+1}$
to the working redex and its reduct, respectively, corresponding to an
$App$- or $\xi$-rule. Notice that the claimed property $\klamp{G_0}\succ\ldots\succ\klamp{G_m}$ for the associated reduction sequence mentioned
above then follows after observing that the algorithm in Definition
\ref{assignalgodefi} commutes with substitution in the sense of 
Lemma \ref{assignsubstlem} and checking the possible cases as outlined in
Definition \ref{correspondencedefi}.

\section{Derivation Lengths Classification}\label{classificationsec}
We are going to show that the upper bounds for reduction sequences in G\"odel's $\GT$ and its fragments
$\GTn$ are optimal. The set of terms $\Tn$ in the fragment $\GTn$, $n\in\Nat$, is the restriction of $\T$  
to recursors of type level $\le n+2$. From our remarks in Subsection \ref{extsubsec} it follows that
we can discuss term reductions in $\T$ using our results regarding reduction sequences in $\Tpr$ via the
mutual embeddings of reduction sequences in $\T$ and $\Tpr$.  
 
\begin{defi} For $G\in\T$ let $\Lv(G)$ denote the maximum type level of subterms of $G$, and let 
$\Rv(G)$ denote the maximum type level of recursors occurring in $G$. We further define
\begin{eqnarray*}
\DT(m)&:=&\max\{k\mid\ex{G_1,\ldots,G_k\in\T} G_1\rhd\ldots\rhd G_k\andsp\lh(G_1),\Lv(G_1)\le m\}\\[1mm]
\DTn(m)&:=&\max\{k\mid\ex{G_1,\ldots,G_k\in\Tn} G_1\rhd\ldots\rhd
G_k\andsp\lh(G_1),\Lv(G_1)\le m\}.\rlap{\hbox to29 pt{\hfill\Fin}}
\end{eqnarray*}\vspace{-9 pt}
\end{defi}
\noindent We are going to use the following common notation for exponential expressions. We set $\om_0:=1$ and $\om_{i+1}:=\om^{\om_i}$,
$2_0(\al):=\al$, and $2_{i+1}(\al):= 2^{2_i(\al)}$ where $\al<\epsn$.

\begin{cor}\label{cortwo} Corollary \ref{assigncor} gives rise to the following derivation lengths classifications.
\begin{enumerate}[\em(1)]
\item The functions definable in $\GT$, i.e.\ the provably recursive
      functions of $\mathrm{PA}$, comprise the $<\!\epsn$-recursive functions.
      The derivation lengths function $\DT$ is $\epsn$-recursive.
\item The functions definable in $\GTn$, i.e.\ the provably recursive
      functions of $\mathrm{I}\Sigma_{n+1}$, comprise the $<\!\om_{n+2}$-recursive functions.
      The derivation lengths function $\DTn$ is $\om_{n+2}$-recursive.
\end{enumerate}
\end{cor}

\noindent\proof
We make use of the well-known fact that $\mathrm{PA}$ has a
functional interpretation in $\GT$ (see \cite{G58,Sh67,S77}) and the fact that
the fragments $\mathrm{I}\Sigma_{n+1}$ have functional interpretations in the $\GTn$
(see \cite{P72}). By results from \cite{BCW94} the corollary then
follows from Corollary \ref{assigncor}.
The detailed argumentation is given below. It is based on preparations worked out in the appendix, \ref{cortwopreps},
which we will use in the form of citations of Lemma \ref{fintwo}, whose purpose
it is to extract bounds on the ordinal vectors $\klamp{G}$ assigned to terms $G$ of $\T$ which are expressed in
terms of maximum type level $\Lv(G)$ and length $\lh(G)$ of $G$.
Notice that we only need to consider the (unique) canonical assignment for the terms of $\T$.

Let $C^0\in\T$ be closed. There is an $m\in\Nat$ such that
\[C\rhdal\ldots\rhdal\underline{m}:\equiv\Suc^{(m)}\N.\] 
We define 
\[\val(C):=m.\]
Theorem \ref{assignthm} shows that $\klampn{C}$ is an upper bound for $\val(C)$ and 
the length of any reduction sequence starting from $C$, since we have
\[m<\klampn{\underline{m}}.\]
Note that
\[\klampn{\underline{m}}\le\psi(\om\cdot(m+1)+m+2)<\psi(\om^2+m)\]
using Proposition \ref{proptwo}, parts \ref{proptwotwo} and \ref{proptwothree}.

Let $F^{00}\in\Tn$ be closed. $F$ represents the function \[m\mapsto \val(F\underline{m}),\]
and for any $m\in\Nat$ we have 
\[\DTF(m),\val(F\underline{m})\le\klampn{F\underline{m}}\]
where
\[
\DTF(m):=\max\{k\mid\ex{G_1,\ldots,G_k\in\T} F\underline{m}\equiv G_1\rhd\ldots\rhd G_k\}.
\]
We have 
\[\klampn{F\underline{m}} =
  \psi(\om\cdot\klampe{F}+\klampn{F}+\klampn{\underline{m}}) <
  \psi(\om\cdot(\klampe{F}+\om^2)+\klampn{F}+m),\]
and the function
\[m\mapsto \psi(\underbrace{\om\cdot(\klampe{F}+\om^2)+\klampn{F}}_
  {<\om_{n+2}\mbox{ \footnotesize by Lemma \ref{fintwo}}}+m)\]
is $<\!\om_{n+2}$-recursive (cf.\ \cite{BCW94}),
implying that also $\DTF$ and
$m\mapsto \val(F\underline{m})$ are
$<\!\om_{n+2}$-recursive. We therefore obtain that the functions definable in $\GTn$
are $<\!\om_{n+2}$-recursive and the functions definable in $\GT$
are $<\!\epsn$-recursive.

Now let some $m\in\Nat$ and a term $G^\si\in\T$ with $\lh(G),\,\Lv(G)\le m$ and
$\Rv(G)\le n+2$ be given. Let $\vecg$ be the closed canonical assignment to the term $G$. 
Then according to Lemma \ref{fintwo}
\begin{eqnarray*}
  g_0 & < &
  \psi(\underbrace{\om\cdot2_{n+1}(\om\cdot
                   2_{m+1}(2(m+1+\lh(G))))}_{<\om_{n+2}})\\[1mm]
  & < & \psi(\om_{n+2}+m)\quad\mbox{by Proposition \ref{proptwo}, part \ref{proptwothree}}.
\end{eqnarray*}
This implies
\[\DTn(m)<\psi(\om_{n+2}+m),\]
and hence $\DTn$ is an $\om_{n+2}$-recursive function.
Omitting the restriction concerning $\Rv(G)$ it similarly follows that
$\DT$ is an $\epsn$-recursive function.
\qed

\section*{Acknowledgements}
The authors would like to thank Roger Hindley and Jonathan Seldin for informative discussions as well as Pierre-Louis Curien for constructive assistance in the rewrite process of the original contribution to TLCA 2009. We are indebted to one of the referees who pointed
out our sloppiness regarding the treatment of bound variables in an ealier version of the
paper and gave many constructive suggestions that have led to an improvement of the first
section of this paper.
We wish to thank William Howard for a number of helpful suggestions and simplifications for the final version of the paper.

\appendix
\section{Proofs omitted in Sections \ref{ordinalsec} and \ref{classificationsec}}

\subsection{Proofs in Subsection \ref{ordtermsubsec}}
We give the detailed proofs of the two Propositions regarding the $\psi$-function.\medskip

\noindent{\bf Proof of Proposition \ref{propone}.}\hfill

\noindent{\bf Ad \ref{proponei}:} This is an immediate consequence of the definitions of
$\oplus$ and $\no$.

\medskip
\noindent{\bf Ad \ref{proponeii}:} Suppose
\begin{eqnarray*}
  \al & = & \om^{\al_1}+\ldots+\om^{\al_n}>\al_1\ge\ldots\ge\al_n
    \mbox{ and}\\[1mm]
  \be & = & \om^{\be_1}+\ldots+\om^{\be_m}>\be_1\ge\ldots\ge\be_m
\end{eqnarray*}
where $n,\,m\ge1$. By definition of $\otimes$ we have
\begin{eqnarray*}
  \no(\al\otimes\be) & = &
    nm + m\sum_{i=1}^n\no(\al_i) + n\sum_{j=1}^m\no(\be_j)\\[1mm]
  & \le &
    nm + m\sum_{i=1}^n\no(\al_i) + n\sum_{j=1}^m\no(\be_j) +
    \sum_{i=1}^n\no(\al_i) \cdot \sum_{j=1}^m\no(\be_j)\\[1mm]
  & = &
    \no(\al) \cdot \no(\be).
\end{eqnarray*}
This estimation of the norm of the natural product holds for
all $\al$ and $\be$. Equality holds if and only if  $\al<\om$ or
$\be<\om$. We further have
\begin{eqnarray*}
  \no(\al)+\no(\be)-1 & = &
    n+m-1 + \sum_{i=1}^n\no(\al_i) + \sum_{j=1}^m\no(\be_j)\\[1mm]
  & \le &
    nm + m\sum_{i=1}^n\no(\al_i) + n\sum_{j=1}^m\no(\be_j)\\[1mm]
  & = &
    \no(\al\otimes\be).
\end{eqnarray*}

\medskip
\noindent{\bf Ad \ref{proponeiii}:} Since the case $\al<\om$ is trivial we may assume that
$\al\ge\om$, say,  $\al=\om\cdot\al_0+m$ where $0<\al_0\le\al$ and $m<\om$.
Let
\[\al_0=\om^{\al_1}+\ldots+\om^{\al_k}+\ldots+\om^{\al_n}
  > \al_1\ge\ldots\ge\al_k\ge\om>\al_{k+1}\ge\ldots\ge\al_n\]
with $0\le k\le n>0$.
Then we have
\[\om\cdot\al_0 = \om^{\al_1}+\ldots+\om^{\al_k}+\om^{\al_{k+1}+1}+\ldots+
  \om^{\al_n+1},\]
which implies $\no(\al)=\no(\al_0)+n-k+m$.
By definition $2^\al=\om^{\al_0}\cdot2^m$, hence
\[\no(2^\al) = (\no(\al_0)+1)\cdot2^m.\]
We obtain from these preparations
\[\no(\al)\le2\no(\al_0)+m<2(\no(\al_0)\cdot2^m+2^m)=2\cdot\no(2^\al)\]
as well as
\[\no(2^\al)=(\no(\al_0)+1)\cdot2^m\le2^{\no(\al_0)}\cdot2^m\le2^{\no(\al)}.\]
This concludes the proof of Proposition \ref{propone}.
\qed\medskip

\noindent{\bf Proof of Proposition \ref{proptwo}.}
The proof of part \ref{proptwoone} is by straightforward induction on $k$ and follows
directly from the definition of $\psi$. Also part \ref{proptwothree} immediately follows from the
Definition of $\psi$. For part \ref{proptwofour} note that $\Phi(\no(\al))=\psi(\Phi(\no(\al)))$ by part \ref{proptwoone},
then apply part \ref{proptwothree}. 
For $\al = 0$ part \ref{proptwotwo} follows immediately from part \ref{proptwoone}. In the case $\al > 0$
the first $\le$-relation is immediate by part \ref{proptwoone}, for the second $\le$-relation we argue by induction on $\be$:
\begin{iteMize}{$\bullet$}
\item $\be = 0$: Trivial.
\item $\be > 0$: By definition of $\psi$ there exists a $\ga<\be$
  such that $\no(\ga)\le\Phi(\no(\be))$ and
  $\psi(\be) = \psi(\ga)+1$. Therefore
  \[\psi(\al+\psi(\be))\le
    \psi(\al\oplus\ga+1)\le
    \psi(\al\oplus\be)\]
  where the first $\le$-relation follows from the i.h.\ for $\al+1$ and the second $\le$-relation is verified using part \ref{proptwothree}. 
  If $\ga+1=\be$ we are done. Otherwise
  we have $\al\oplus\ga+1<\al\oplus\be$ and
  \[\no(\al\oplus\ga+1)\le\no(\al)+1+\Phi(\no(\be))\le
    \Phi(\no(\al\oplus\be)),\]
  using that $\al>0$.
  \qed
\end{iteMize}

\subsection{Proofs in Subsection \ref{boxsubsec}}
We give the detailed proofs of the lemmas regarding the $\bop$-operator.\medskip

\noindent{\bf Proof of Lemma \ref{boxnormlem}.}
The proof is essentially the same as in \cite{W98}. We give the details for the reader's convenience, frequently using Proposition 
\ref{propone}.
We first show the following

\medskip
\noindent{\bf Claim.} For $k\le i\le n$ we have
  \begin{equation}\label{boxnormclaim}
  \no((\vecf\bopd\vecg)_i)\prec F_0^{2(n+1\minusp i)} (f+g+1)
  \end{equation}
where $n\minusp i:=n-i$ if $n\ge i$, and $n\minusp i:=0$ otherwise.
The claim is shown by induction on $n\minusp i$. For $i=n$ we obtain
\begin{eqnarray*}
  \no((\vecf\bopd\vecg)_n)
  & = & \no(2^{f_{n+1}}\cdot(f_n+g_n)) \\[1mm]
  & \preceq & 2^{\no(f_{n+1})}\cdot(\no(f_n) + \no(g_n)) \\[1mm]
  & \preceq & 2^{f}\cdot(f+g) \\[1mm]
  & \prec & 2^{2^{f+g}}\cdot(f+g) \\[1mm]
  & \prec & 2^{2^{f+g+1}} \\[1mm]
  & = & F_0^2(f+g+1).
  \end{eqnarray*}
For $k\le i<n$ we have 
\begin{eqnarray*}
  \no((\vecf\bopd\vecg)_i)
  & = & \no(2^{(\vecf\bopd\vecg)_{i+1}}\cdot(f_i+g_i)) \\[1mm]
  & \preceq & 2^{F_0^{2(n+1-i)-2}(f+g+1)}\cdot(f+g) \\[1mm]
  & = & F_0^{2(n+1-i)-1}(f+g+1)\cdot(f+g) \\[1mm]
  & \prec & F_0^{2(n+1-i)}(f+g+1).
  \end{eqnarray*}
Now we prove the lemma. For $k,n+1\le i\le m$ we have
\[\no((\vecf\bopd\vecg)_i) = \no(f_i) \preceq f \prec F_2(f+g+n).\]
For $k\le i\le n$ we obtain 
\begin{eqnarray*}
  \no((\vecf\bopd\vecg)_i)
  & \prec & F_0^{2(n+1\minusp i)} (f+g+1) \mbox{ by \ref{boxnormclaim}}\\[1mm]
  & \preceq & F_0^{2n}(f+g+1) \\[1mm]
  & \preceq & F_1(f+g+2n) \\[1mm]
  & \prec & F_2(f+g+n),
\end{eqnarray*}
concluding the proof of the lemma.
\qed\medskip

\noindent{\bf Proof of Lemma \ref{boxthree}.}
Part \ref{boxthreei} is adapted from \cite{H70}, p.\ 450, and is shown here for the reader's 
convenience.

\noindent{\bf Ad \ref{boxthreei}:} We argue by induction on $n+1\minusp i$ for $1\le i\le n$. 
  \begin{eqnarray*}
    \lefteqn{(\veca\bopd\vecd)_i+(\vecb\bopd\vecd)_i} \\[1mm]
    &\prec&
     2^{(\veca\bopd\vecd)_{i+1}}\cdot(a_i+b_i+d_i)
     +
     2^{(\vecb\bopd\vecd)_{i+1}}\cdot(a_i+b_i+d_i)\\[1mm]
    &\preceq&
     2^{(\veca\bopd\vecd)_{i+1}+(\vecb\bopd\vecd)_{i+1}}\cdot
     (a_i+b_i+d_i)\\[1mm]
    &\preceq&
     2^{(\vecc\bopd\vecd)_{i+1}}\cdot(c_i+d_i) \\[1mm]
    &=&
     (\vecc\bopd\vecd)_i.
  \end{eqnarray*}

\medskip
\noindent{\bf Ad \ref{boxthreeii}:} For $1\le i\le m$ we set 
  \[N_i:= \sum^{m}_{j=i}\no(a_j),\quad
    M_i:= \sum^{m}_{j=i}\no(b_j),\quad 
    L_i:= \sum^{n}_{j=i}\no(d_j).\]
Lemma \ref{boxnormlem} yields for $1\le i\le m$
\begin{equation}\label{boxthreeiieqone}
\no((\veca\bopd\vecd)_i)\preceq F_2(N_i+L_i+n).
\end{equation}
Since the case $i > n$ is trivial we assume in the sequel that
$1\le i \le n$. 
By Proposition \ref{propone}, part \ref{proponeiii}, we have 
\[\no((\vecb\bopd\vecd)_{i+1})\preceq2\no((\vecb\bopd\vecd)_i).\]
From this we obtain for $i\le j\le n+1$
\[\no((\vecb\bopd\vecd)_j)\preceq2^n\cdot\no((\vecb\bopd\vecd)_i).\]
Since $\no(b_j)\preceq \no((\vecb\bopd\vecd)_j)$ for every $j$  and
$\no(d_j)\preceq \no((\vecb\bopd\vecd)_j)$ for  $1\le j\le n$
we get
\begin{equation}\label{boxthreeiieqtwo}
M_i+L_i \preceq (n+1)\cdot 2^{n+1}\cdot \no((\vecb\bopd\vecd)_i).
\end{equation}
\ref{boxthreeiieqone} and \ref{boxthreeiieqtwo} together with $N_i\preceq M_i$ yield
\begin{eqnarray*}
\no((\veca\bopd\vecd)_i) & \preceq & F_2(M_i+L_i+n)\\[1mm]
  & \preceq & F_2((n+1)\cdot2^{n+1}\cdot \no((\vecb\bopd\vecd)_i)+n)\\[1mm]
  & \preceq & F_3(\no((\vecb\bopd\vecd)_i)+n).
\end{eqnarray*}

\medskip
\noindent{\bf Ad \ref{boxthreeiii}:} For $n<i\le m$ the claim holds by assumption, and for
$1\le i\le n$ the claim follows by part \ref{boxthreei}.
Consider the case $i=0$ (the inequalities are explained below):
\begin{eqnarray*}
\lefteqn{(\veca\bopd\vecd)_0+(\vecb\bopd\vecd)_0} \\[1mm]
& = &
    \psiterm{(\veca\bopd \vecd)_1}{ a_0+ d_0+n} +
    \psiterm{(\vecb\bopd \vecd)_1}{ b_0+ d_0+n} \\[1mm]
& \preceq &
\psiterm{((\veca\bopd\vecd)_1+(\vecb\bopd\vecd)_1)}
        { a_0+ b_0+2 d_0+2n} \\[1mm]
& \prec &
\psiterm{(\vecc\bopd\vecd)_1}{ c_0+ d_0+n}\\[1mm]
& = & 
(\vecc\bopd\vecd)_0.
\end{eqnarray*}
The $\preceq$-relationship follows by Proposition \ref{proptwo}, parts \ref{proptwoone} and \ref{proptwotwo}.
We will now show that the $\prec$-relationship holds, making use of Proposition \ref{proptwo}, part \ref{proptwothree}, and 
part \ref{boxthreeii} of the present lemma:
\[(\veca\bopd\vecd)_1+(\vecb\bopd\vecd)_1\prec(\vecc\bopd\vecd)_1\]
holds by assumption if $n = 0$, and for $n>0$ this has already been shown.
We obtain
\[\om\cdot((\veca\bopd \vecd)_1+(\vecb\bopd \vecd)_1)
  + a_0+ b_0+2 d_0+2n \prec
  \om\cdot(\vecc\bopd \vecd)_1+ c_0+ d_0+n.\]
Part \ref{boxthreeii} yields
\[\no((\veca\bopd\vecd)_1), \no((\vecb\bopd\vecd)_1) \preceq
  F_3(\no(\vecc\bopd\vecd)_1+n).\]
From this we easily verify the second assumption of Proposition \ref{proptwo}, part \ref{proptwothree}.
\qed\medskip

\noindent{\bf Proof of Lemma \ref{boxfour}.}
We proceed by induction on $n+1\minusp i$.
If $i = n+1$, then
  \[2e_i = 2a_i\prec c_i
   = (\vecc\bopd\vecd)_i.\]
In the case $i = n$ we have
  \begin{eqnarray*}
    e_n
    &=&
      2^{a_{n+1}}\cdot (2^{a_{n+1}}\cdot(a_n+d_n)
      +2^{b_{n+1}}\cdot(b_n+d_n))\\[1mm]
    &\prec&
      2^{2a_{n+1}}
        \cdot(a_n+b_n+d_n)
      +
      2^{a_{n+1}+b_{n+1}}
        \cdot(a_n+b_n+d_n)\\[1mm]
    &\preceq&
      2^{2a_{n+1}+b_{n+1}}
        \cdot(a_n+b_n+d_n).
  \end{eqnarray*}
  This implies
  \[2e_n \prec
    2^{c_{n+1}}\cdot(c_n+d_n) = (\vecc\bopd\vecd)_n.\]
Now let us assume that $1\le i<n$:
  \begin{eqnarray*}
    e_i
    &=&
      2^{e_{i+1}}
      \cdot 
      (2^{(\veca\bopd\vecd)_{i+1}}\cdot(a_i+d_i)
        +2^{(\vecb\bopd\vecd)_{i+1}}\cdot(b_i+d_i))\\[2mm]
    &\prec&
      2^{e_{i+1}+(\veca\bopd\vecd)_{i+1}}
        \cdot(a_i+b_i+d_i)
      + 
      2^{e_{i+1}+(\vecb\bopd\vecd)_{i+1}}
        \cdot(a_i+b_i+d_i)\\[1mm]
    &\preceq&
      2^{2e_{i+1}}
      \cdot
      (c_i+d_i)
  \end{eqnarray*}
  where the last $\preceq$-relation holds since $i+1\le n$ and by Lemma \ref{boxone}
  \[(\veca\bopd\vecd)_{i+1}+(\vecb\bopd\vecd)_{i+1} \preceq
     e_{i+1}.\]
  Using the i.h., which allows us to estimate $2e_i\prec 2^{2e_{i+1}+1}\cdot(c_i+d_i)\preceq2^{(\vecc\bopd\vecd)_{i+1}}\cdot(c_i+d_i)$, 
  we finally obtain
  \[2e_i\prec
    (\vecc\bopd\vecd)_i.\]
This concludes the proof of the lemma.
\qed\medskip

\noindent{\bf Proof of Lemma \ref{boxextnormlem}.}
For convenience we set $a:=a_0$, $b:=b_0$, $c:=c_0$, and $d:=d_0$.
We first show the following

\medskip
\noindent{\bf Claim.} For $1\le i\le n$, setting $e:=a+b+c+d+2(n+1)$, we have
\begin{equation}\label{boxextnormclaim}
\no(e_i)\preceq F_2^{2(n+1\minusp i)}(e).
\end{equation}
The claim is proved by induction on $n\minusp i$.
We will make use of the following abbreviations:
\begin{eqnarray*}
  \al & := & F_2(a+b+n), \\[1mm]
  \be & := & F_2(c+d+n), \\[1mm]
  \ga & := & F_2(a+b+c+d+2n+1).
\end{eqnarray*}
If $i = n$, then we obtain
\begin{eqnarray*}
  \no(e_n)
  & = & \no(2^{a_{n+1}}\cdot
           ((\veca\bopd\vecb)_n+(\vecc\bopd\vecd)_n)) \\[1mm]
  & \preceq & 2^{\no(a_{n+1})}\cdot
              (\no((\veca\bopd\vecb)_n)+\no((\vecc\bopd\vecd)_n)) \\[1mm]
  & \preceq & 2^a\cdot(\al+\be) \mbox{ by Lemma \ref{boxnormlem}}\\[1mm]
  & \preceq & F_2(\ga+1) \\[1mm]
  & \preceq & F_2^2(e).
\end{eqnarray*}
If $1\le i<n$, then we have
\begin{eqnarray*}
  \no(e_i)
  & \preceq & 2^{\no(e_{i+1})}
              \cdot(\no((\veca\bopd\vecb)_i)+\no((\vecc\bopd\vecd)_i)) \\[1mm]
  & \preceq & 2^{F_2^{2(n+1\minusp i)-2}(e )}\cdot(\al+\be) \mbox{ by Lemma \ref{boxnormlem} and the i.h.}\\[1mm]
  & \preceq & F_2^{2(n+1\minusp i)-1}(e)\cdot\ga \\[1mm]
  & \prec & F_2^{2(n+1\minusp i)}(e ).
\end{eqnarray*}
Now we prove the lemma from the above claim.
The case $i=n+1$ is trivial since we then have $e_{n+1}=a_{n+1}$. 
For $1\le i\le n$ we finally obtain 
\begin{eqnarray*}
  \no(e_i)
  & \preceq & F_2^{2(n+1\minusp i)}(e ) \mbox{ by \ref{boxextnormclaim}}\\[1mm]
  & \preceq & F_2^{2n}(e ) \\[1mm]
  & \prec & F_3(a+b+c+d+n).
\end{eqnarray*}
This concludes the proof of Lemma \ref{boxextnormlem}.
\qed

\subsection{Proofs in Subsection \ref{dopsec}}
We provide the proofs regarding the operator $\dop$.
Our first goal is to show that the operators $\dopx$ preserve bounded norms. To this end, and in preparation of the analysis of our assignment in Section \ref{classificationsec}, we need to introduce precise notions of subterms.

\begin{defi}\label{subxdefi}
By recursion on the buildup of $h\in\Cxi$ we define
the set $\txij(h)$ of maximal $x$-free subterms of $j$-th level of $h$
and the set $\subxij(h)$ of those subterms of $j$-th level of
$h$ which are different from $x_i$, where $x$-free subterms are considered atomic. 
We use the abbreviation
\[\hij:=\{h\:\mid\:i=j\}.\]
If $h$ is $x$-free, then \[\txij(h) := \hij =: \subxij(h),\]
otherwise we distinguish between the following cases:
      \begin{iteMize}{$\bullet$}
      \item If $h \equiv x_i$, then
              \[\txij(h) := \emptyset =: \subxij(h).\]
      \item If $h \equiv f + g$, then
              \begin{eqnarray*}
              \txij(h)&:=& \txij(f) \cup \txij(g),\\[2mm] 
              \subxij(h)&:=& \hij\cup\subxij(f)\cup\subxij(g).
              \end{eqnarray*}
      \item If $h \equiv 2^f\cdot g$ or $h \equiv \psiterm{f}{g}$, then
              \begin{eqnarray*}
              \txij(h)&:=&\txipj(f) \cup \txij(g),\\[2mm] 
              \subxij(h)&:=& \hij\cup\subxipj(f)\cup\subxij(g).
              \end{eqnarray*}
      \end{iteMize}
We further define \[\txi(h) := \bigcup_{i \le j} \txij(h)\quad\mbox{ and }\quad\subxi(h) := \bigcup_{i \le j} \subxij(h).\]\Fin
\end{defi}
Notice that for $h\in \Cxi$ the set $\txij(h)$ comprises the $x$-free terms of $\subxij(h)$,
that we have $\txij(h) = \subxij(h) = \emptyset$ if $i>j$, and that
for every $t \in \subxii(h)$ we have $t \preceq h$.

\begin{lem}\label{doplem} Let $\vecx$ be a variable vector and set $k:=\lv(\vecx)$,
$n:=k+1$. 
\begin{enumerate}[\em(1)]
\item\label{doplemone} For any $h\in\Cxi$ and $t\in\subxi(h)$ we have
\[\no(t)\preceq 2^{n\minusp i}\cdot\no(h).\]
\item\label{doplemtwo} Let $\vech\in\Cx$ be of bounded norm, 
$m:=\lv(\vech)$.
Then for all $t\in\txi(h_i)$, $i\le m$, we have 
\begin{eqnarray}\label{doplemtwoi}
\no(t)\prec 2^{n\minusp i}\cdot(\dopxsin h_0)_0,
\end{eqnarray}
and for $0<j\le n$ we have
\begin{eqnarray}\label{doplemtwoii}
\no((\dopxsii h_i)_j)\preceq\lhx(h_i)\cdot2^n\cdot(\dopxsin h_0)_0.
\end{eqnarray}
\item\label{doplemthree} 
$\dopx$ preserves bounded norm: for every $\vech\in\Cx$ of bounded
norm $\dopx\vech$ is of bounded norm.
\end{enumerate}
\end{lem}
\proof Part \ref{doplemone} is shown by induction on the buildup of $h$.
If $i>k$, then $h$ is $x$-free, so $t\equiv h$ and we are done.
Suppose $i\le k$. If $h$ is of a form $\psiterm{f}{g}$, we use Proposition
\ref{proptwo} to see that $\no(f),g\preceq h$.
In the interesting case, where $h\equiv2^f\cdot g$ and $t\in\subxip(f)$,
we have $\no(t)\preceq 2^{n\minusp(i+1)}\cdot\no(f)$, and by Proposition
\ref{propone} we have 
$\no(f)\preceq2\no(2^f)\preceq 2\no(h)$.
Therefore, $\no(t)\preceq 2^{n\minusp i}\cdot\no(h)$.

We turn to the proof of part \ref{doplemtwo}. By part \ref{doplemone} we
have $\no(t)\preceq 2^{n\minusp i}\cdot\no(h_i)$. Since $t$ is $x$-free
and $\vech$ of bounded norm, we even have 
\[\no(t)\preceq 2^{n\minusp i}\cdot\no(h_i\{\vecx:=\vec{1}\})
\preceq 2^{n\minusp i}\cdot h_0\{\vecx:=\vec{1}\},\]
which by Lemma \ref{crucialestimlem} implies \ref{doplemtwoi}.
In order to show \ref{doplemtwoii}, set 
$\nu:=2^n\cdot(\dopxsin h_0)_0$. We show by induction on the buildup of
$h\in\Cxi$ that if $\no(t)\prec\nu$ for all $t\in\txi(h)$, then
\[\no((\dopxsii h)_j)\preceq\lhx(h)\cdot\nu.\]
If $h$ is $x$-free, we obtain 
$\no((\dopxsii h)_j)\preceq\no(h)+1\preceq\nu$. Now suppose $h$ is 
not $x$-free. The case $h\equiv x_i$ is trivial, the case $h\equiv f+g$ 
follows directly from the i.h., and in the remaining cases notice
that $\txip(f),\txi(g)\ssq\txi(h)$. For $h\equiv2^f\cdot g$ we obtain
\[\no((\dopxsii h)_j)\preceq2\no((\dopxsiip f)_j)+\no((\dopxsii g)_j)+1,\]
which implies the claimed estimate since $\lhx(h)=2\lhx(f)+\lhx(g)+1$.
The remaining situation $h\equiv\psiterm{f}{g}$ is handled 
similarly.

Part \ref{doplemthree} is now easy to see. In the case $j>n$ we apply
Lemma \ref{crucialestimlem} to obtain
\[\no((\dopx\vech)_j)=\no(h_j)\preceq h_0\{\vecx:=\vec{1}\}
\prec(\dopxsin h_0)_0\preceq(\dopx\vech)_0.\] 
For $0<j\le n$ we apply \ref{doplemtwoii} to obtain
\[\no((\dopx\vech)_j)=\sum^m_{i=0}\no((\dopxsii h_i)_j)\preceq
\left(\sum^m_{i=0}\lhx(h_i)\right)\cdot 2^n\cdot(\dopxsin h_0)_0
=(\dopx\vech)_0.\]
Thus $\dopx\vech$ is of bounded norm. 
\qed

\noindent{\bf Proof of Lemma \ref{interplaylem}.}
For convenience we set $k:=\lv(\vecx)$, $n:=k+1$, $m:=\lv(\vech)$, 
and $l:=\max\{m,n\}$.
We first show that the lemma is a consequence of the following 
\medskip

\noindent{\bf Claim.} For $h\in\Cxi$, $i\le n$, we have
\begin{equation}\label{interplayclaim}
h\prec(\dopxsii h\bopd\vecx)_i.
\end{equation}
In order to derive the lemma from \ref{interplayclaim}, let $i\le m$. We distinguish between the following three cases.

\medskip 
{\bf Case 1:} $n<i\le l$. Then clearly
   $h_i\equiv(\dopx\vech\bopd\vecx)_i$.

\medskip 
{\bf Case 2:} $1\le i\le n$. We have 
      \[(\dopxsii h_i)_j\preceq(\dopx\vech)_j\]
      for $i\le j\le n$. Thus by Lemma \ref{boxtwo}, part \ref{boxtwoi},
      \[(\dopxsii h_i\bopd\vecx)_i \preceq (\dopx\vech\bopd\vecx)_i,\]
      as we may ignore components of $\dopx\vech$ above the $n$-th.
      By \ref{interplayclaim} we have
      $h_i\prec(\dopxsii h_i\bopd\vecx)_i$.

\medskip 
{\bf Case 3:} $i=0$. Here \ref{interplayclaim} applies since we have
   \begin{eqnarray*}
      (\dopxsin h_0 \bopd\vecx)_0 &=&  
      \psiterm{(\dopxsin h_0 \bopd\vecx)_1}
              {(\dopxsin h_0 )_0+x_0+k}\\[1mm]
      & \prec &        
      \psiterm{(\dopx\vech\bopd\vecx)_1}
              {(\dopx\vech)_0+x_0+k} \\[1mm]
      & = & (\dopx\vech\bopd\vecx)_0
   \end{eqnarray*}
   by Proposition \ref{proptwo}, part \ref{proptwothree}, whose assumptions are
   easily checked: For all $j\le n$ we have
         $(\dopxsin h_0 )_j \prec (\dopx\vech)_j$. Lemma \ref{boxtwo}, part \ref{boxtwoi}, yields
         $(\dopxsin h_0 \bopd\vecx)_1\prec(\dopx\vech\bopd\vecx)_1$.
         By Lemma \ref{boxthree}, part \ref{boxthreeii}, $\no((\dopxsin h_0 \bopd\vecx)_1)\preceq
         F_3(\no((\dopx\vech\bopd\vecx)_1)+k)$.

\medskip 
We now prove Claim \ref{interplayclaim} by induction on the definition of $\dopxsii$.
Assume first that $h$ is $x$-free.
Then clearly $h\prec h+1=(\dopxsii h)_i\preceq(\dopxsii h\bopd\vecx)_i$.
Otherwise we must have $i\le k$ and distinguish between the following four cases:

\medskip 
{\bf Case 1:} $h\equiv x_i$. $h\prec x_i+1\preceq(\vec{1}\bopd\vecx)_i$.

\medskip 
{\bf Case 2:} $h\equiv f+g$. Then we apply the i.h.\,and use Lemma \ref{boxthree}, part \ref{boxthreeiii}:
      \begin{eqnarray*}
        h\equiv f+g
        & \prec &
        (\dopxsii f\bopd \vecx)_i + (\dopxsii g\bopd \vecx)_i \\[1mm]
        & \prec &
        (\dopxsii h\bopd\vecx)_i.
      \end{eqnarray*}

\medskip 
{\bf Case 3:} $h\equiv 2^f\cdot g$. Then $i\ge1$, and after applying the i.h.\,we use Lemma \ref{boxfour}:
      \begin{eqnarray*}
        h\equiv 2^f\cdot g & \prec &
        2^{(\dopxsiip f\bopd \vecx)_{i+1}}\cdot(\dopxsii g\bopd \vecx)_i \\[1mm]
        & \preceq &
        ((\dopxsiip f\bopd \vecx) \bopd
        (\dopxsii g\bopd \vecx\!\restriction_k))_i \\[1mm]
        & \prec &
        (\dopxsii h\bopd\vecx)_i.
      \end{eqnarray*}

\medskip 
{\bf Case 4:} $h\equiv\psiterm{f}{g}$. Then we have $i=0$ and obtain
      \begin{eqnarray*}
        h
        & \prec &
        \psi(\om\cdot(\dopxsie f\bopd\vecx)_1+
            (\dopxsin g\bopd\vecx)_0) \mbox{ (see below)}\\
        & = &
        \psi(\om\cdot(\dopxsie f\bopd\vecx)_1+
             \psi(\om\cdot(\dopxsin g\bopd\vecx)_1
             +(\dopxsin g)_0+ x_0 +k)) \\
        & \preceq &
        \psi(\om\cdot((\dopxsie f\bopd\vecx)_1+(\dopxsin g\bopd\vecx)_1)+
                (\dopxsin g)_0+ x_0 +k) \\
        & \preceq &
        \psi(\om\cdot(\dopxsin h\bopd\vecx)_1+
                (\dopxsin h)_0+ x_0 +k) \mbox{ (see below)} \\
        & = &
        (\dopxsin h\bopd\vecx)_0.
      \end{eqnarray*}
      The strict inequality follows from the i.h., using that
     \[\no(f)\preceq F_2(g) \prec F_2((\dopxsin g\bopd\vecx)_0).\]
     The last inequality is easily verified: 
     In the case $k>0$ Lemma \ref{boxthree}, part \ref{boxthreei}, yields 
     \[(\dopxsie f\bopd\vecx)_1+(\dopxsin g\bopd\vecx)_1 \prec(\dopxsin h\bopd\vecx)_1,\] 
     and for $k=0$ both terms are equal.
     Using Lemma \ref{boxthree}, part \ref{boxthreeii}, we obtain
     \[\no((\dopxsie f\bopd\vecx)_1),
        \no((\dopxsin g\bopd\vecx)_1)
        \preceq F_3(\no((\dopxsin h\bopd\vecx)_1)+k).\]
     Clearly, we have $(\dopxsin g)_0\preceq(\dopxsin h)_0$.
\qed

\subsection{Preparations for the proof of Corollary \ref{cortwo} in Section \ref{classificationsec}}\label{cortwopreps}
Here we are going to show Lemmata \ref{finone} and \ref{fintwo}, where the former will be used in the proof of the latter.
As a preparation for the proof of Lemma \ref{finone}, recall the definition of $\lhx$ in Definition \ref{dopxdefi}. 
We define the variable independent version $\sz$ to 
serve as another auxiliary function in order to estimate the term complexity of ordinal terms occurring in our assignment.
Note that we have $\lhx\le\sz$.
\begin{defi}
$\sz(h)$ for $h\in\ot$ is defined by
\begin{iteMize}{$\bullet$}
\item $\sz(h):=1$ if $h$ is a variable or constant.
\item $\sz(h):=\sz(f)+\sz(g)+1$ if $h$ is of a form either $f+g$ or $\psiterm{f}{g}$.
\item $\sz(h):=2\sz(f)+\sz(g)+1$ if $h$ is of a form $2^f\cdot g$. \Fin
\end{iteMize}
\end{defi}

\begin{lem}\label{szlem} Suppose $h\in\Cxi$ for some $i\in\Nat$. We have
\[\sz((\dopxsii h)_j)\le4\sz(h)\]
for $j\le\lv(\vecx)+1$.
\end{lem}
\proof The proof is by straightforward induction on the buildup of $h$, along the Definition of $\dopxsii h$.
\qed

\begin{lem}\label{finone}
Let $G\in\T$ and $\vecg:=\klamp{G}$ be its canonical assignment. 
Setting $m:=\lv(\vecg)$ we have
\begin{equation}\label{finoneclaimone}
\sum^m_{i=0}\sz(g_i)<2_2(\Lv(G)+2\lh(G))=:M(G).
\end{equation}
Setting $n:=\lv(\vecx)+1$ and
$L_n(G):=\max\{n,\Lv(G)\}$ we have
\begin{equation}\label{finoneclaimtwo}
\lx(\vecg)<2_2(L_n(G)+1+2\lh(G)). 
\end{equation}
\end{lem}
\proof The proof of \ref{finoneclaimone} is by induction on $\lh(G)$.
The second claim \ref{finoneclaimtwo} then follows from the first one since $n\le L_n(G)$.

\medskip 
{\bf Case 1:} $G$ is a constant or variable. Then the claim is trivial once
we notice that $m\le \Lv(G)$.

\medskip 
{\bf Case 2:} $G\equiv AB$. Let $\veca$ and $\vecb$ be the canonical assignments of $A$ and $B$, respectively, and set $m_A:=\lv(\veca)$,
$m_B:=\lv(\vecb)$. The vector $\vecc:=\veca\bopd\vecb$ agrees with 
$\vecg$ up to component $m$, and for $m_B<i\le m_A$ we have $c_i=a_i$, 
hence by the i.h.
\[\sz(c_i)=\sz(a_i)<M(A).\]
By side induction on $m_B+1\minusp i$ we show
\begin{equation}\label{finoneappclaim}
\sz(c_i)<2^{2(m_B+1\minusp i)}(M(A)+M(B))
\end{equation}
The case $i=m_B+1$ has already been taken care of.
Suppose $i\le m_B$. If $i>0$ we have 
$\sz(c_i)=2\sz(c_{i+1})+\sz(a_i)+\sz(b_i)+2$,
while $\sz(c_i)=\sz(c_{i+1})+\sz(a_i)+\sz(b_i)+2(m_B+1)$ for $i=0$.
In any case we obtain
\begin{eqnarray*}
\sz(c_i)
&<&2^{2(m_B\minusp i)+1}(M(A)+M(B))+M(A)+M(B)\\[1mm]
&<&2^{2(m_B+1\minusp i)}(M(A)+M(B)).
\end{eqnarray*}
Using \ref{finoneappclaim} the formula for the geometric series yields  
\[\sum^{m_B}_{i=0}\sz(c_i)<4^{m_B+2}(M(A)+M(B)),\]
hence, together with the i.h.\ applied to $A$ 
\begin{eqnarray*}
\sum^m_{i=0}\sz(g_i)&<&(4^{\Lv(G)+1}+1)(M(A)+M(B))\\[1mm]
&<&2_2(\Lv(G)+2\lh(G)-1)
   \cdot(2_2(\Lv(G)+2\lh(A))+2_2(\Lv(G)+2\lh(B)))\\[1mm]
&\le&(2_2(\Lv(G)+2\lh(G)-1))^2\\[1mm]
&=&M(G)
\end{eqnarray*}
showing \ref{finoneappclaim} for application terms.

\medskip 
{\bf Case 3:} $G\equiv\lam Y.F$. Let $\vecf$ be the canonical assignment
to $F$, set $k:=\lv(\vecf)$, and $l:=\lv(\vecy)+1$. We have 
$\vecg=\dopy\vecf+\vecf\{\vecy:=\vec{1}\}$ and $m=\max\{k,l\}$. 
Setting \[M:=2_2(\Lv(G)+2\lh(F)),\] 
by the i.h., \ref{finoneclaimtwo}, applied to $F$ we have
\begin{eqnarray*}
\ly(\vecf)&<&
2_2(L_l(F)+1+2\lh(F))\\[1mm]
&\le&2_2(\Lv(G)+1+2\lh(F))\\[1mm]
&=&M^2
\end{eqnarray*}
since $L_l(F)=\max\{l,\Lv(F)\}\le \Lv(G)$, and by the i.h., \ref{finoneclaimone}, we have
\[\sum^k_{j=0}\sz(f_j)<M(F)\le M,\] 
whence using Lemma \ref{szlem} we obtain
\[\sz(g_i)<\left\{\begin{array}{l@{\mbox{ if }}l}
M^2(4\sz(f_0)+1)&i=0\\[1mm]
4M+k+\sz(f_i)+2&1\le i\le l\\[1mm]
2\sz(f_i)+2&l<i\le m.
\end{array}\right.\]
We may now generously estimate the sum of the above terms:
\begin{eqnarray*}
\sum^m_{i=0}\sz(g_i)&<&
M^2(4\sz(f_0)+1)+(4\Lv(G)+2)M+(\Lv(G)+2)^2\\[1mm]
&<&M^2(4\sz(f_0)+1)+(4\Lv(G)+3)M\\[1mm]
&<&M^2(4\sz(f_0)+2)\\[1mm]
&<&M^4\\[1mm]
&=&M(G)
\end{eqnarray*} 
which concludes the proof of Lemma \ref{finone}.
\qed

The following two lemmas will prepare the proof of Lemma \ref{fintwo}. Recall Definition \ref{subxdefi}, and
for any $h\in\ot$ let $\bar{h}$ be the closure of $h$ by replacing every variable in $h$ with $1$.

\begin{lem}\label{subydopxlem}
Suppose $h\in\Cxk$ and $i<j$ where $i\le n:=\lv(\vecx)+1$. Then we have
\[\subyij((\dopxsik h)_i)\ssq\subykj(h\{\vecx:=\vec{1}\}).\]
\end{lem}
\proof The proof is by induction on the buildup of $h$ along the definition of the partial operators $\dopxsik$, $k\in\Nat$.
If $h$ is $x$-free, then we have $(\dopxsii h)_i=h+1$ if $k=i$ while $(\dopxsik h)_i=1$ if $k\not= i$, and the claim follows 
immediately. Now suppose that $h$ is not $x$-free. We then distinguish between the following cases.  

{\bf Case 1:} $h$ is a variable or constant. Since $i<j$ we then have $\subyij((\dopxsik h)_i)=\emptyset$, so there
is nothing to show.

{\bf Case 2:} $h\equiv f+g$. Then we have $(\dopxsik h)_i=(\dopxsik f)_i+(\dopxsik g)_i+1$, and therefore
\[\subyij((\dopxsik h)_i)=\subyij((\dopxsik f)_i)\cup\subyij((\dopxsik g)_i)\] since $i<j$. Thus we may directly apply the i.h.

{\bf Case 3:} $h\equiv2^f\cdot g$. Then $(\dopxsik h)_i=2(\dopxsikp f)_i+(\dopxsik g)_i+1$, and we argue as in the previous case.

{\bf Case 4:} $h\equiv\psi(\om\cdot f+g)$. Here the case $k\not= i$ is treated as the previous cases, so assume $k=i=0$,
whence $(\dopxsin h)_0=\psi(\om\cdot f\{\vecx:=\vec{1}\}+(\dopxsin g)_0)$. We therefore have
\[\subynj((\dopxsin h)_0)=\subyej(f\{\vecx:=\vec{1}\})\cup\subynj((\dopxsik g)_i)\] since $j>0$, and clearly
\[\subyej(f\{\vecx:=\vec{1}\})\cup\subynj(g\{\vecx:=\vec{1}\})=\subynj(h\{\vecx:=\vec{1}\}),\]
applying the i.h.\ for $g$ if necessary.
\qed

\begin{lem}\label{dopxboundslem}\mbox{ }
\begin{enumerate}[\em(1)]
\item\label{dopxboundslemone} Let $h\in\Cxi$, $n:=\lv(\vecx)+1$, $j\in(0,n]$, and $\al\in(0,\epsn)$ such that
$\bar{t}<\al$ for all $t\in\txij(h)$. Then we have
\[\overline{(\dopxsii h)_j}\le\lhx(h)\cdot\al.\]
\item\label{dopxboundslemtwo} Let $h\in\Cxn$, $m\in(0,\om)$, and $\al\in(0,\epsn)$ such that
$\bar{t}<m$ for all $t\in\txnn(h)$ and $\bar{f}<\al$ for all $f\in\subxne(h)$. Then we have
\[\overline{(\dopxsin h)_0}<\psi(\om\cdot\lhx(h)\cdot\al+\lhx(h)\cdot m).\]
\end{enumerate}  
\end{lem}
\proof Part \ref{dopxboundslemone} is proved by induction on the buildup of $h\in\Cxi$.
If $h$ is $x$-free or $h\equiv x_i$, the claim follows immediately. Let us assume that $h$ is not $x$-free.
In the case $h\equiv f+g$ the claim directly follows from the i.h.\ for $f$ and $g$. 
If $h\equiv2^f\cdot g$ or $h\equiv\psiterm{f}{g}$, we have $\txij(h)=\txipj(f)\cup\txij(g)$, 
and straightforwardly apply the i.h.\ to $f$ and $g$.

Part \ref{dopxboundslemtwo} is shown by induction on the buildup of $h\in\Cxn$ along the definition of $\dopxsin h$.
If $h$ is $x$-free, then $h\in\txnn(h)$, and we have
  \[\overline{(\dopxsin h)_0} = \overline{h}+1 \le m < \psi(\om\cdot\lhx(h)\cdot\al+\lhx(h)\cdot m).\]
Let us now assume that $h$ is not $x$-free. 
The case $h\equiv\xsinull$ is trivial.
If $h\equiv f+ g$, then using Proposition \ref{proptwo} and the i.h.\ we may estimate straightforwardly as follows:
    \begin{eqnarray*}
      \overline{(\dopxsin h)_0} 
      & = &
        \overline{(\dopxsin f)_0}+\overline{(\dopxsin g)_0}+1\\[1mm]
      & < &
        \psi(\om \cdot \lhx(f) \cdot \al+\lhx(f) \cdot  m) +
        \psi(\om \cdot \lhx(g) \cdot \al+\lhx(g) \cdot  m) \\[1mm]
      & \le &
        \psi(\om\cdot(\lhx(f)+\lhx(g))\cdot\al
             +(\lhx(f)+\lhx(g))\cdot m) \\[1mm]
      & < &
        \psi(\om\cdot\lhx(h)\cdot\al+\lhx(h)\cdot m).
    \end{eqnarray*}
Finally, suppose $h\equiv\psiterm{f}{g}$. Since $f\in\subxne(h)$ we have
    $\overline{f}<\al$, and since $h\in\Cxn$ we have $\no(f)\preceq F_2(g)$.
    By Lemma \ref{crucialestimlem} we obtain
    \[\no\left(\overline{f}\right)\le F_2\left(\overline{g}\right) < F_2\left(\overline{(\dopxsin g)_0}\right),\]
    and using the i.h., Proposition \ref{proptwo} yields
    \begin{eqnarray*}
      \overline{(\dopxsin h)_0} & = &
      \psi\left(\om\cdot\overline{f}+\overline{(\dopxsin g)_0}\right)\\[1mm]
      & < &
        \psiterm{\al}{\psi(\om\cdot\lhx(g)\cdot\al+\lhx(g)\cdot m)}\\[2mm]
      & \le &
        \psi(\om\cdot(\lhx(g)+1)\cdot\al+(\lhx(g)+1)\cdot m)\\[2mm]
      & \le &
        \psi(\om\cdot\lhx(h)\cdot\al+\lhx(h)\cdot m),
    \end{eqnarray*}
concluding the proof of Lemma \ref{dopxboundslem}.
\qed

\begin{lem}\label{fintwo}
Let $G\in\T$, $L:=\Lv(G)$, $R:=\Rv(G)$, and for every subterm $H$ of $G$ set $M_H:=2(L+1+\lh(H))$
and define a vector $\vecal^H$ of level $L$ by
\[\al_i^H:=\left\{\begin{array}{l@{\mbox{ if }}l}
2_{L+2\minusp i}(M_H\minusp i)& R<i\le L,\\[1mm]
2_{R\minusp i}(\om\cdot2_{L+2\minusp i}(M_H\minusp i))& 1\le i\le R,\\[1mm]
\psi(\om\cdot2_{R\minusp 1}(\om\cdot2_{L+1}(M_H)))& i=0.
\end{array}\right.
\]
Then for every subterm $H$ of $G$ with canonical assignment 
$\vech:=\klamp{H}$ we have
\begin{equation}\label{fintwoclaimone}
\bar{h}_i<\al_i^H
\end{equation}
for $i\le \lv(\vech)=:m$.
If $\vech\in\Cy$, then for all $i\le m$, all $j\le L$, and every $t\in\subyij(h_i)$ we have
\begin{equation}\label{fintwoclaimtwo}
\bar{t}<\al_j^H.
\end{equation}
\end{lem}
\proof The lemma is proved by induction on $\lh(H)$. 

{\bf Case 1:} $H$ is a variable or constant. Then the claims follow immediately.
The subcase $H\equiv\R$ is where infinite ordinals enter the picture.

{\bf Case 2:} $H\equiv A^{\si\tau}B^\si$. Then $\vech=\veca\bopd\vecb\!\restriction_m$ where
$\veca$ and $\vecb$ are the canonical assignments to $A$ and $B$, respectively. Let $n:=\lv(\si)$.
We first show \ref{fintwoclaimone}.
If $i>n$ we have $\bar{h}_i\le \bar{a}_i$, and the claim follows by the i.h.\ applied to $A$. Suppose $i\le n$, whence
$i<L$. We argue by side induction on $n\minusp i$.
  \begin{iteMize}{$\bullet$}
  \item $L >i>R $: Then $0<i\le n$ and hence
    \begin{eqnarray*}
      \bar{h}_i & \le & 2^{\bar{h}_{i+1}}\cdot(\bar{a}_i+ \bar{b}_i)\\[1mm]
      & < &
        2_{L +2\minusp i}(M_H\minusp(i+1))\cdot
        (2_{L +2\minusp i}(M_A\minusp i)+2_{L +2\minusp i}(M_B\minusp i))\\[2mm]
      & < &
        (2_{L +2\minusp i}(M_H\minusp(i+1)))^2\\[2mm]
      & < &
        2_{L +2\minusp i}(M_H\minusp i).
    \end{eqnarray*}
  \item $R \ge i\ge 1$: Suppose first that $i=R $. Then we have 
    \begin{eqnarray*}
      \bar{h}_i & < &
        2_{L +2\minusp i}(M_H\minusp(i+1))\cdot
        (\om\cdot2_{L +2\minusp i}(M_A\minusp i)+
         \om\cdot2_{L +2\minusp i}(M_B\minusp i))\\[2mm]
      & < &
        2_{L +2\minusp i}(M_H\minusp(i+1))\cdot
        (\om\cdot2_{L +2\minusp i}(M_H\minusp (i+1)))\\[2mm]
      & < &
        \om\cdot2_{L +2\minusp i}(M_H\minusp i).
    \end{eqnarray*}
    We now consider the case $i<R $ where we estimate as follows.
    \begin{eqnarray*}
      \bar{h}_i & < &
        2_{R \minusp i}(\om\cdot2_{L +2\minusp (i+1)}
        (M_H\minusp(i+1)))\cdot\\
        & & \qquad
          (2_{R \minusp i}(\om\cdot2_{L +2\minusp i}
          (M_A\minusp i))+
           2_{R \minusp i}(\om\cdot2_{L +2\minusp i}
           (M_B\minusp i)))\\[2mm]
        & < &
          2_{R \minusp i}(\om\cdot2_{L +2\minusp (i+1)}
          (M_H\minusp(i+1)))\cdot\\
        & & \qquad
          2_{R \minusp i}(\om\cdot2_{L +2\minusp i}
          (M_H\minusp(i+1)))\\[2mm]
        & < &
          2_{R \minusp i}(\om\cdot2_{L +2\minusp i}
          (M_H\minusp i)).
    \end{eqnarray*}
  \item $i=0$: In case of $R =0$ we obtain using Proposition \ref{proptwo}
    \begin{eqnarray*}
      \bar{h}_0 & < &
        \psi(\om\cdot2_{L +1}(M_H\minusp 1)+
        \psi(\om^2\cdot2_{L +1}(M_A))+
        \psi(\om^2\cdot2_{L +1}(M_B))+n)\\[2mm]
      & < &
        \psi(\om^2\cdot2_{L +1}(M_H)).
    \end{eqnarray*}
    If $R >0$ we even have $R \ge 2$, and using again Proposition \ref{proptwo} we obtain
    \begin{eqnarray*}
      \bar{h}_0 & < &
        \psi(\om\cdot2_{R \minusp1}
          (\om\cdot2_{L +1}(M_H\minusp1))+\\
        & & \qquad  
        \psi(\om\cdot2_{R \minusp1}(\om\cdot2_{L +1}(M_A)))+
        \psi(\om\cdot2_{R \minusp1}(\om\cdot2_{L +1}(M_B)))+n)\\[2mm]
      & < &
        \psi(\om\cdot2_{R \minusp1}(\om\cdot2_{L +1}(M_H))).
    \end{eqnarray*}
  \end{iteMize}
  This finishes the verification of \ref{fintwoclaimone}, and we proceed with proving \ref{fintwoclaimtwo}. 
  If $\vech\in\Cy$, then we must also have $\veca,\vecb\in\Cy$ and may apply the respective i.h.'s. 
  Suppose $t\in\subyij(h_i)$, whence $i\le j$, as the set $\subyij(h_i)$ is empty if $i>j$. 
  In order to show $\bar{t}<\al^H_j$ we employ an induction on $j\minusp i$. 
  If $i=j$ we clearly have $\bar{t}\le\bar{h_i}<\al^H_i$ by \ref{fintwoclaimone}.
  Suppose $i<j$. In the case $i>n$ we use the i.h.\ applied to $A$. 
  If on the other hand $i\le n$, then we have
  \[t\in\subyipj(h_{i+1})\cup\subyij(a_i)\cup\subyij(b_i),\]
  and in each case $\bar{t}<\al^H_j$ follows using the i.h.\ since clearly $\al^A_j,\al^B_j<\al^H_j$.

{\bf Case 3:} $H\equiv \lam X^\si\!.F^\tau$. Then $\vech=\dopx\vecf+\vecf\{\vecx:=\vec{1}\}$ 
where $\vecf:=\klamp{f}$, and $m=\max\{n,l\}$ where $n:=\lv(\vecx)+1$ and $l=\lv(\vecf)$.
We first show \ref{fintwoclaimone}, where we distinguish between the following three cases.
\begin{iteMize}{$\bullet$}
\item $n<i\le m$. Then we have $h_i= 2f_i$, and the claim follows easily from the i.h.\ for $F$.
\item $1\le i\le n$. By part \ref{dopxboundslemone} of Lemma \ref{dopxboundslem} and the i.h.\ for $F$ we have
\[
\overline{(\dopxsik f_k)_i}\le\lhx(f_k)\cdot \al^F_i
\]
for every $k\le l$, which yields 
\[\bar{h}_i\le\left(\sum^l_{k=0}\lhx(f_k)+1\right)\cdot\al^F_i\le2_2(L+2\lh(F))\cdot\al^F_i\]
by \ref{finoneclaimone} of Lemma \ref{finone}.
    In case of $i>R$ we may now estimate
    \begin{eqnarray*}
      \bar{h}_i & \le &
        2_2(L +2\lh(F))\cdot2_{L +2\minusp i}(M_F\minusp i)\\[2mm]
      & < &
        (2_{L +2\minusp i}(M_F\minusp i))^2\\[2mm]
      & < &
        2_{L +2\minusp i}(M_H\minusp i)\\[2mm]
      & = &
        \al^H_i,
    \end{eqnarray*}
    whereas in case of $i\le R$ we estimate
    \begin{eqnarray*}
      \bar{h}_i & \le &
        2_2(L +2\lh(F))\cdot
        2_{R\minusp i}(\om\cdot2_{L +2\minusp i}
          (M_F\minusp i))\\[2mm]
      & < &
        2_{R\minusp i}(\om\cdot2_{L +2\minusp i}
          (M_H\minusp (i+1)))\\[2mm]
      & < &
        \al^H_i.
    \end{eqnarray*}
\item $i=0$. Since in the case $R=0$ the argumentation is easier, let us assume that $R>0$.
By Lemma \ref{finone} we have
\[\lhx(f_0), \lx(\vecf)<2_2(L+1+2\lh(F))=:K,\]
and relying on the i.h.\ for $F$ we obtain using part \ref{dopxboundslemtwo} of Lemma \ref{dopxboundslem}
\[
\overline{(\dopxsin f_0)_0}<\psi(\om\cdot\lhx(f_0)\cdot\al^F_1+\lhx(f_0)\cdot\al^F_0),
\]
which, using Proposition \ref{proptwo}, allows for the following estimation:
    \begin{eqnarray*}
      \bar{h}_0 & = &
        \lx(\vecf)\cdot\overline{(\dopxsin f_0)_0}+
        \bar{f}_0\\[2mm]
      & < &
        K\cdot\psi(\om\cdot K\cdot \al^F_1+ K\al^F_0)\\[2mm]
      & \le &
        \psi(\om\cdot K^2\cdot \al^F_1+K^2\al^F_0)\\[2mm]
      & < &
        \psi(\om\cdot2K^2\cdot
        2_{R\minusp1}(\om\cdot2_{L +1}(M_F)))\\[2mm]
      & < &
        \psi(\om\cdot2_{R\minusp1}(\om\cdot2_{L +1}(M_H)))\\[2mm]
      & = &
        \al^H_0.
    \end{eqnarray*}
\end{iteMize}

In order to verify \ref{fintwoclaimtwo} suppose $t\in\subyij(h_i)$. In the case $i=j$ we obtain
$\bar{t}\le\bar{h}_i<\al^H_i$ from \ref{fintwoclaimone} by the monotonicity properties of $+,\cdot,$
and $\psi$. Assume $i<j$. Then we either have $t\in\subyij(f_i\{\vecx:=\vec{1}\})$ where $i\le l$, or $i\le n$ and
\[t\in\subyij((\dopxsik f_k)_i)\] for some $k\le l$, which is $0$ in the case $i=0$, and  
Lemma \ref{subydopxlem} yields $t\in\subykj(f_k\{\vecx:=\vec{1}\})$, hence $k\le j$.
If $\vecy\equiv\vecx$, we must have $k=j$, $t\equiv f_j\{\vecx:=\vec{1}\}$, and therefore $\bar{t}\le\al^F_j$ by the i.h.\ for $F$.
Now assume $\vecy\not\equiv\vecx$. Then we have
\[\subykj(f_k\{\vecx:=\vec{1}\})=\subykj(f_k)\{\vecx:=\vec{1}\},\]
and $\bar{t}\le\al^F_j$ follows from the i.h.\ for $F$. The case $t\in\subyij(f_i\{\vecx:=\vec{1}\})$ is treated 
in the same way.
\qed

\end{document}